\documentclass[graphics, twocolumn, usenatbib]{mn2e}
\usepackage{natbib} 
\usepackage{epsfig} 
\usepackage{graphicx} 
\usepackage{color}
\usepackage{aas_macros} 
\usepackage{amssymb}
\usepackage{amsmath}
\usepackage[title]{appendix}

\newcommand{\enzo}{\texttt{Enzo~}}

\newcommand{\msolar} {$\rm{M_{\odot}}~$}
\newcommand{\msolarc} {$\rm{M_{\odot}}$}
\newcommand{\molH} {$\rm{H_2}$~}

\newcommand{\J} {$\rm{10^{-21}\ erg\ cm^{-2}\ s^{-1}\ Hz^{-1}\ sr^{-1}}$}

\newcommand{\JU} {$\rm{ erg\ cm^{-2}\ s^{-1}\ Hz^{-1}\ sr^{-1}}$}

\def\etal{{\it et al.}~}
\begin{document}
\title[]{The effect of dark matter resolution on the collapse of baryons in high 
  redshift numerical simulations}

\author[J.A. Regan \etal] 
{John A. Regan$^{1}$\thanks{E-mail:john.regan@helsinki.fi}, Peter H. Johansson$^{1}$ 
\& John H. Wise$^{2}$ \\ \\
$^1$Department of Physics, University of Helsinki, Gustaf H\"allstr\"omin katu 2a,
FI-00014 Helsinki, Finland \\
$^2$Center for Relativistic Astrophysics, Georgia Institute of Technology, 837 State Street, 
Atlanta, GA 30332, USA
\\}


\maketitle

\begin{abstract}
We examine the impact of dark matter particle resolution on the formation of a 
baryonic core in high resolution adaptive mesh refinement simulations. We test the 
effect that both particle smoothing and particle splitting have on the hydrodynamic 
properties of a collapsing halo at high redshift ($z > 20$). Furthermore, we vary the 
background field intensity, with energy below the Lyman limit ($< 13.6$ eV), as may be 
relevant for the case of metal-free star formation and super-massive black hole seed
formation. We find that using particle splitting methods greatly increases 
our particle resolution without introducing any numerical noise and allows us to achieve
converged results over a wide range of external background fields. Additionally, we find that
for lower values of the background field a lower dark matter particle mass is required. 
We define the radius of the core as the point at which the enclosed baryonic mass 
dominates over the enclosed dark matter mass. For our simulations this results in  
$\rm{R_{core} \sim 5\ pc}$. We find that in order to produce converged results which are 
not affected by dark matter particles requires that the relationship  
${M_{\rm{core}} / M_{\rm{DM}}} > 100.0$ be satisfied, where ${M_{\rm{core}}}$ is the enclosed baryon 
mass within the core and $M_{\rm{DM}}$ is the minimum dark matter particle mass. This ratio 
should provide a very useful starting point for conducting convergence tests before any 
production run simulations. We find that dark matter particle smoothing is a useful adjunct 
to already highly resolved simulations.
\end{abstract}



\section{Introduction} 
 
Gravitational collapse of dark matter has been well studied over the past four decades 
with increasingly high resolution N-body numerical simulations 
\citep{Davis_1985, Frenk_1988, Warren_1992, Cen_1994, Gelb_1994, Navarro_1996, Navarro_1997, 
Hernquist_1996, Jenkins_2001, Wambshanss_2004, Springel_2008, Diemand_2008, Bosch_2014} leading to 
the confirmation that the large-scale structure of the Universe is built up through hierarchical
collapse of small dark matter haloes. It is within the gravitational potential wells of these dark 
matter structures that galaxies and their associated active galactic nuclei and quasars 
form \citep{White_1978, White_1991, Springel_2005a, Croton_2006a, Johansson_2012, 
Vogelsberger_2014}. \\
\indent Dark matter is modelled as a collisionless 
fluid and in principle can be solved using the collisionless Boltzmann equation \citep[e.g.][]
{Shlosman_1979, Rasio_1989, Hozumi_1997}. However, the computational demands of solving 
this six dimensional equation means that instead dark matter is typically treated in a 
Monte-Carlo fashion (however see recent work by \citealt{Hahn_2015})
and represented by a set of collisionless particles which interact with each other gravitationally. 
Solving the gravitational interactions of a many-body system is then dealt with in one of two ways, 
direct summation \citep[e.g.][]{VonHoerner_1960, Aarseth_1963, VonHoerner_1963, Steinmetz_1996} 
or more commonly, in cosmological simulations, using an approximate treatment involving a Tree 
algorithm \citep{Barnes_1986, Barnes_1989} or a particle mesh algorithm 
\citep{Efstathiou_1985, Hockney_1988} - see also \cite{Bertschinger_1998} for a comprehensive 
review. \\
\indent The subsequent addition of gas dynamics to the simulations increases the complexity of 
the simulations substantially, requiring not just a detailed treatment of the gas (fluid) 
interaction with the dark matter, but also of a range of physical processes involving the gas 
including radiative cooling, star formation and the tracking of individual chemical species. 
Incorporating the gas dynamics 
into the simulation usually also takes the form of one of two mechanisms - smoothed particle 
hydrodynamics (SPH) \citep{Lucy_1977, Gingold_Monaghan_1977} or grid based methods 
\citep{Ryu_1990, Cen_1990, Stone_1992, Bryan_1995} where the hydrodynamical equations are solved 
on the grid using finite difference techniques possibly involving deforming or adaptive meshes 
\citep[e.g.][]{Springel_2010, Enzo_2014}. \\
\indent Here we focus on the effect the mass resolution of the dark matter particles can have on the 
gas dynamics within a collapsing halo where the gas densities can exceed the dark matter densities
by several orders of magnitude as happens in the centres of newly forming galaxies, mini-haloes and 
clusters of galaxies. In order to conduct this case study we use the \enzo code (see 
\S \ref{Sec:NumericalSetup} for details on \texttt{Enzo}) to follow the collapse of gas 
within collapsing haloes early in the Universe at a time when the cooling of the gas is driven by 
primordial processes and in which the gas has not yet been enriched with metals from previous 
episodes of star formation. This
case is relevant for simulations involving the first generation of stars (Pop III stars) 
\citep{Abel_2002, Bromm_2002, Yoshida_2003a, Bromm_2004, Yoshida_2006, OShea_2008, 
Yoshida_2008, Wise_2008b, OShea_2008, Turk_2009, Bromm_2009, Clark_2011, Stacy_2010, Stacy_2012,
Susa_2014, Stacy_2014, Hirano_2014} or 
for the case of direct collapse models of supermassive black holes \citep{Shlosman_1989, 
Haiman_2001, Oh_2002, Bromm_2003, Haiman_2006, Begelman_2006, Wise_2008a, Regan_2009b, 
Regan_2009, Regan_2014a, Regan_2014b} which are 
also thought to form out of pristine gas at high redshift. \\
\indent Previous studies of dark matter resolution have indicated the negative numerical effects 
that dark matter can have on the collapse of baryons in the context of galaxy formation. 
\cite{Steinmetz_1997} showed that using dark matter particles with masses above a critical mass 
limit (their work was focused in the context of galaxy formation) introduced spurious two 
body heating effects which they found completely suppressed the effects of radiative cooling. 
Although their numerical tests were carried out using the SPH technique they argued that the 
results are valid in the case of grid based approaches also given the modelling of the dark matter 
component is similar in both instances. And while other studies have noted the effect that dark 
matter mass resolution plays in galaxy formation simulations \citep[e.g.][]{Machacek_2001, 
Kaufmann_2006} no specific study has been made, to the best of our knowledge, on the effect of 
dark matter resolution on the collapsing core\footnote{The core radius is here defined as the 
radius at which the baryonic enclosed mass exceeds the dark matter mass for the first time. 
Typically, for our simulations this occurs between 1 and 10 parsecs from the point of highest 
density. This definition is closely related to the Jeans length, $ \lambda_J $, of the gas for 
a gas with an isothermal temperature of 8000 K and number density, 
$\rm{n \sim 1 \times 10^6\ cm^{-3}}$.}
within an embryonic galaxy. This is particularly 
crucial as correctly accounting for the cooling and heating effect becomes extremely important 
in determining the characteristics of the final object. In this paper we study this exact effect. \\
\indent It is also worth noting that apart from the negative impact that poorly resolved dark 
matter particles may have on the baryonic physics an additional numerical artefact may occur if a 
dark matter particle flies into the high density region from a lower density region. This 
``flyby'' and the perturbing effects of its large mass may then induce incorrect physical 
conditions. While we do not specifically deal with this issue in this study it can be alleviated 
using particle smoothing (see \S \ref{Sec:ParticleSmoothing}) which will necessarily prevent
adverse effects from a large dark matter mass within the baryon dominated core.  \\
\indent We vary the mass resolution of the dark matter particles that surround the 
collapsing core in an effort to identify any spurious heating effects that the dark matter has on
the collapsing core and identify resolution guidelines. Furthermore, we vary the background 
radiation field resulting in the formation (and collapse) of mini-haloes and haloes with virial 
temperatures T $> 10^4$ K and again test for dark matter mass resolution convergence. \\
\indent The paper is laid out as follows: in \S \ref{Sec:NumericalSetup} we describe the 
numerical approach used and the various techniques employed that can affect the dark matter 
resolution; in \S \ref{Sec:Simulations} we describe the characteristics of the simulations used in 
this study;  in \S \ref{Sec:Results} we describe the results of our 
numerical simulations; in \S \ref{Sec:Choosing} we discuss the parameters for choosing the 
correct dark matter particle mass and in \S \ref{Sec:Conclusions} we present our conclusions.  
Throughout this paper we  assume a standard $\Lambda$CDM cosmology with the following parameters 
\cite[based on the latest Planck data]{Planck_2014}, $\Omega_{\Lambda,0}$  = 0.6817, 
$\Omega_{\rm m,0}$ = 0.3183, $\Omega_{\rm b,0}$ = 0.0463, $\sigma_8$ = 0.8347 and $h$ = 0.6704. 
We further assume a spectral index for the primordial density fluctuations of $n=0.9616$.\\

\begin{figure*}
  \centering 
  \begin{minipage}{175mm}      \begin{center}
      \centerline{
        \includegraphics[width=9cm]{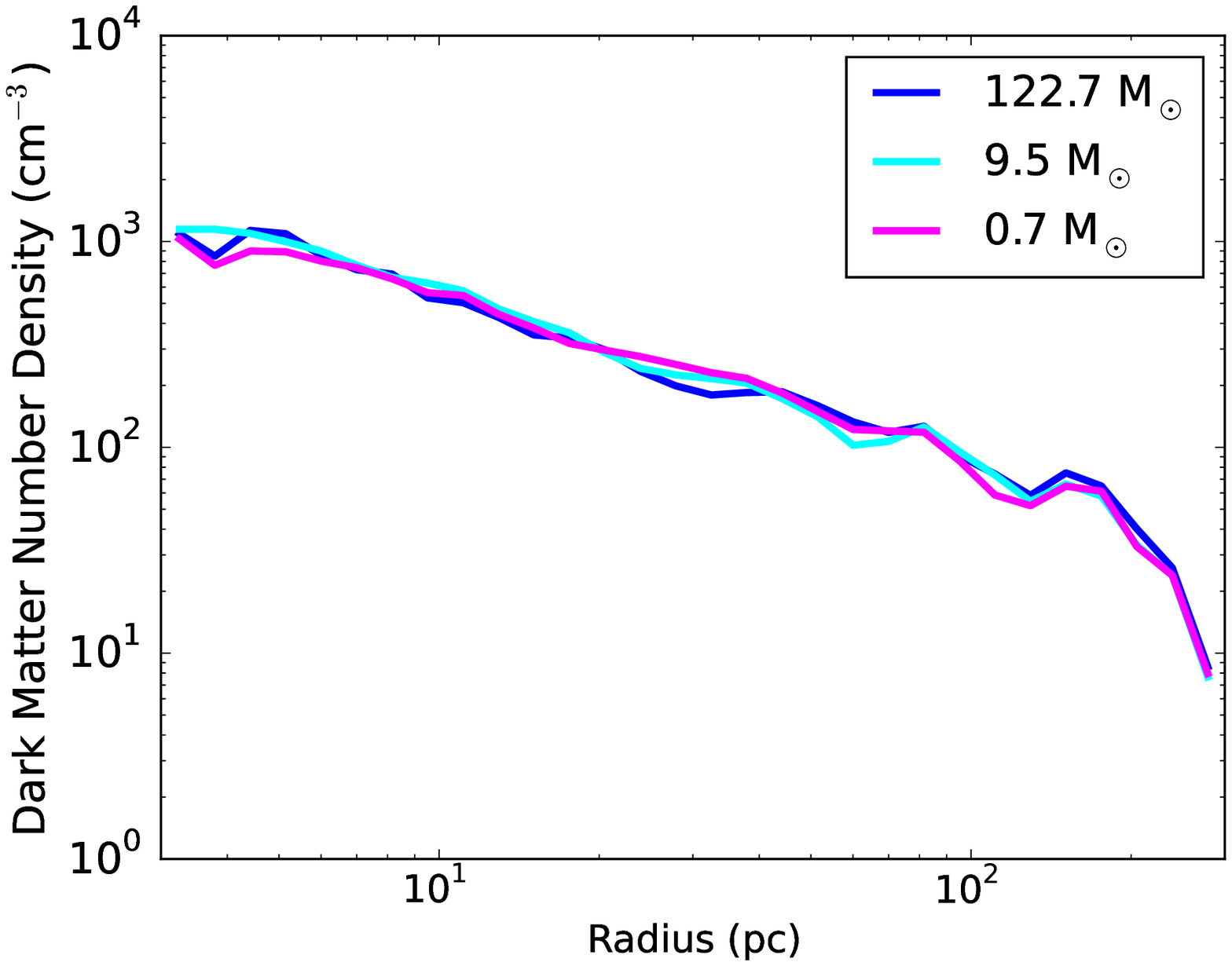}
        \includegraphics[width=9cm]{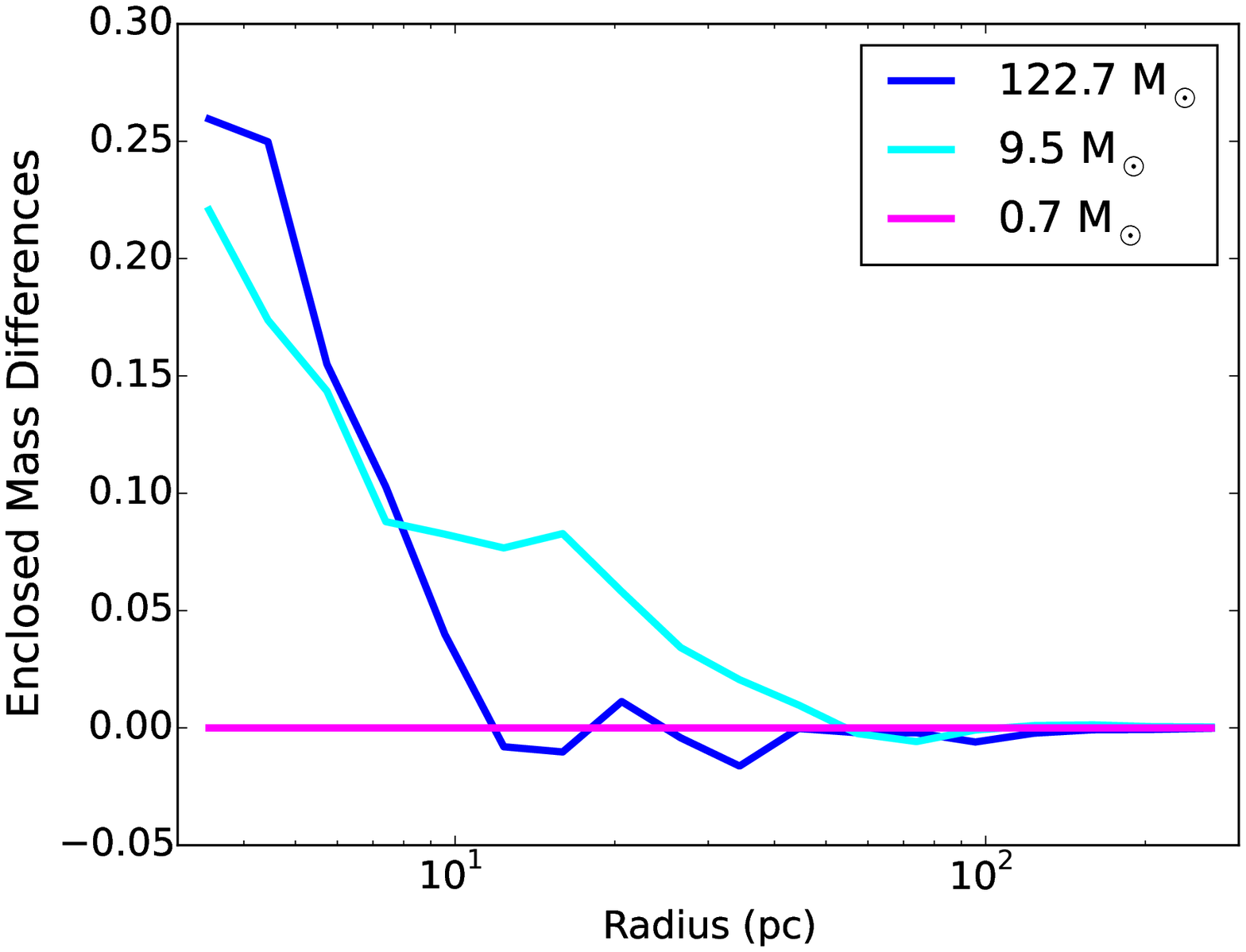}}
        \caption[]
        {\label{Splitting}
          The left hand panel shows the spherically averaged dark matter density
          profile for a dark matter only simulation of Halo A. The output is taken at 
          a redshift of z = 23 and the density is scaled by the mass of the hydrogen atom to 
          enable a direct comparison with the gas density. The lines are for 
          a minimum dark matter mass of 122.7 \msolarc, 9.5 \msolar and 0.7 \msolarc.
          The right hand panel shows the difference between the enclosed mass profile 
          for the same halo at the same output time. The differences are calculated against the 
          highest resolution simulation (i.e. the dark matter simulation with M = 0.7 \msolarc).
          
          }
      \end{center} \end{minipage}
  \end{figure*}

\section{Numerical Setup} \label{Sec:NumericalSetup}
\noindent We have  used the publicly available adaptive mesh refinement
(AMR) code \texttt{Enzo}\footnote{http://enzo-project.org/}. AMR codes
are a popular tool for studying the gravitational and hydrodynamical collapse of 
astrophysical objects  (although see for example \citealt{Yoshida_2006}, 
\citealt{Clark_2011a} \& \citealt{Mayer_2014} among others for collapse simulations conducted 
using SPH techniques) due to their inherent ability to extend over many magnitudes in dynamic range.
Therefore using \enzo as the code with which to perform this study is an obvious choice. 
Furthermore, we use version 3.0\footnote{Changeset: 6037:1c232317e50d} which is
the bleeding edge version of the code incorporating a range of new features (e.g. Goldbaum et al. 
in prep). \\
\indent \enzo was originally developed by Greg Bryan and Mike Norman at the University of 
Illinois \cite[]{Bryan_1995b, Bryan_1997, Norman_1999, OShea_2004, Enzo_2014}. The gravity solver 
in \enzo uses an $N$-Body adaptive particle-mesh technique \citep{Efstathiou_1985,  
Hockney_1988, Couchman_1991} while the hydrodynamics are evolved using the piecewise parabolic 
method combined with a non-linear Riemann solver for shock capturing. The AMR methodology allows for 
additional finer meshes to be laid down as the simulation runs to enhance the resolution
in a given, user defined, region. \\
\indent For our simulations we set the  maximum refinement level to 18. Refinement is triggered
in \enzo  when the refinement criteria are exceeded. The refinement criteria used in this work 
were based on three physical measurements: (1) The dark matter particle over-density, 
(2) The baryon over-density and (3) the Jeans length. The first two criteria introduce additional 
meshes when the over-density (${\Delta \rho \over \rho_{\rm{mean}}}$) of a grid cell with respect to 
the mean density exceeds 8.0 for baryons and/or DM. Furthermore, we set the 
\emph{MinimumMassForRefinementExponent} parameter to $-0.1$ making the simulation super-Lagrangian 
and therefore reducing the threshold for refinement as higher densities are reached. For the final 
criteria we set the number of cells per Jeans length to be 16 in these runs.

\subsection{Chemical Modelling}
We adopt a nine species model for the chemical 
network used including ${\rm H}, {\rm H}^+, {\rm He}, {\rm He}^+,  {\rm He}^{++}, {\rm e}^-, 
\rm{H_2}, \rm{H_2^+}\ and\ \rm{H^-}$. The gas is allowed to cool radiatively during the collapse. 
Additionally, we incorporate an external radiation field with an energy below the hydrogen 
ionisation edge into the simulations conducted here. Specifically we incorporate radiation energies
from 0.76 eV up to 13.6 eV which affects the three species $\rm{H^{-}, H_2, H_2^+}$ as described in 
more detail below. The intensity of the background radiation 
is varied over a range of values between 0 and $500 \times 10^{-21}$ \JU \ with the effective 
temperature of the blackbody temperature set to T = 50000 K in all cases. We use the variable 
J$_{21}$ as shorthand for $10^{-21}$ \JU. We initially concentrate our study on radiation fields 
with 80 J$_{21}$ and  500 J$_{21}$ and then in \S \ref{Sec:Fext} we explore 
the effect of varying the external field.  The shape of the 
spectrum is fully described by the temperature of the blackbody radiation, T$_{\rm{eff}}$ and the 
amplitude of the fluctuation, $\kappa$ J$_{21}$, where $\kappa$ is any scalar quantity. We normalise 
the external radiation field at the hydrogen ionisation edge. In our simulations we allow the 
external field to affect the rates of three species - $\rm{H^{-}, H_2, H_2^+}$, the rates for the 
three species are given below \citep{Abel_1997}:
\begin{equation}
\begin{split}
  k_{H^-} & = {4 \pi \over h} \int_{0.76}^{13.6} J_{21} {B(E, T_{eff}) \over B(13.6, T_{eff})}{\sigma(E) \over E} dE \ \ \ [\rm{s^{-1}}]\\ \\
  k_{H_2} & = 1.38 \times 10^9 J_{21}  {B(E_{LW}, T_{eff}) \over B(13.6, T_{eff})} \ \ \ \ \ \ \ \ \ \ \ \  [\rm{s^{-1}}] \\ \\
  k_{H_2^+} & = {4 \pi \over h} \int_{2.65}^{13.6} J_{21} {B(E, T_{eff}) \over B(13.6, T_{eff})}{\sigma(E) \over E} dE \ \ \ [\rm{s^{-1}}]\\
\end{split}
\end{equation}
where k$_X$ is the photoionisation rate for a given species, J$_{21}$ is the external field with 
implied units of \J, $\sigma(E)$ is the cross section at a given energy (frequency) 
and B(E, T$_{eff}$) is the blackbody spectrum at an energy, E, temperature, T$_{eff}$ and 
$\rm{E_{LW}}$ is the energy in the Lyman-Werner band (12.8 eV). The 
integration limits are in eV and do not exceed the hydrogen ionisation threshold of 13.6 eV.

\subsection {Realisations}
Using the initial conditions generator \texttt{MUSIC} \citep{Hahn_2011} we generated over 
5000 random dark matter realisations with \enzo using a $256^3$ grid. Each simulation was run until 
a redshift of z = 30 with an inline ``Friends of Friends'' halo finder. Several realisations were 
identified as having a high sigma peak ($\sigma > 4.0$ at z $> 30$, see for example 
\citealt{Regan_2014a} for details on the definition of $\sigma$). Three realisations were 
found from the initial dark matter runs and a single halo realisation was used to conduct this 
study - labeled \textit{Halo A}. It should be noted that 
we have also conducted resolution tests on the two other haloes found - \textit{Halo B} and 
\textit{Halo C} - however the results were broadly similar. We discuss the 
characteristics of \textit{Halo B} and \textit{Halo C} in \S \ref{Sec:CompareBC} but focus the 
bulk of this study around Halo A.\\
\indent Each realisation was then resimulated with baryons included and with different levels 
of nested grids and hence dark matter particle resolution. We ran simulations utilising 
one, two and three levels of grid nesting. Each nested grid reduces the dark 
matter particle mass resolution by a factor of eight. Memory limitations restrict the number of 
nested grids to a maximum of three. For this study we use a box size of 2.0 h$^{-1}$ Mpc which 
allows us to find large haloes (high sigma peaks) while also restricting the dynamic range 
required by the simulation. This means that the maximum dark matter particle resolution that \enzo
can achieve in the simulation is $M_{\rm{DM}} \sim 103$ \msolarc. \enzo does not dynamically 
refine the particle mass during the simulation when refining its grids. In order to 
refine the dark matter particles further we employ a Particle Splitting procedure. 
\begin{figure*}
  \centering 
  \begin{minipage}{175mm}      \begin{center}
      \centerline{
        \includegraphics[width=18cm]{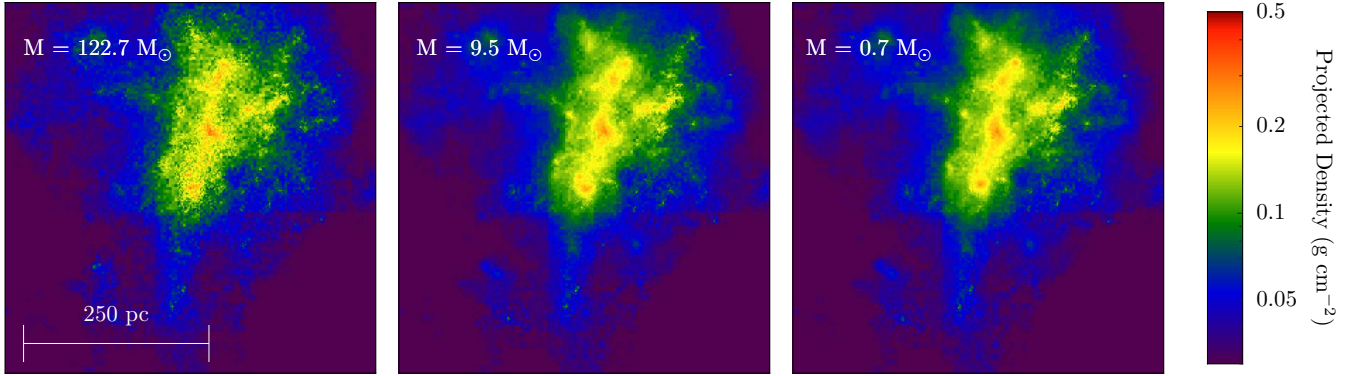}}
        \caption[]
        {\label{Splitting_Visualisation}
          Each panel shows a dark matter density projection of the 0.5 kpc (physical) 
          surrounding the collapsing halo at a redshift of $z = 23$. The left hand 
          panel has a minimum dark matter mass of $M_{\rm{DM}} \sim 122.7$ \msolarc,
          the middle panel has a minimum dark matter mass of $M_{\rm{DM}} \sim 9.5$ \msolarc,
          and the right hand panel has a minimum dark matter mass of $M_{\rm{DM}} \sim 0.7$ \msolarc.
          Each projection is centred on the point of maximum density in the simulation 
          with a dark matter particle mass of $M_{\rm{DM}} \sim 122.7$ \msolarc.
          
          }
      \end{center} \end{minipage}
  \end{figure*}

\begin{table*}
\centering
\caption{Halo Details}
\begin{tabular}{ | l | l | l | l | l | l | l }
\hline \hline 
\textbf{\em $\rm{Halo\ Name}^{a}$}
& \textbf{\em $\rm{M_{Tot}}^{b}$} & \textbf{\em $\rm{M_{\rm{DM}}}^{c}$}
& \textbf{\em $\rm{M_{Baryon}}^{d}$} & \textbf{\em $R_{200}^{e}$} 
& \textbf{\em $T_{\rm vir}^{f}$} & \textbf{\em $n_{\rm H, \rm max}^{g}$} \\
\hline 
Halo A &  $2.3 \times 10^{7}$ & $2.0 \times 10^{7}$ & $3.0 \times 10^{6}$ & 0.35 & 9718 
& $1.05 \times 10^9$ \\
Halo B & $2.3 \times 10^{7}$ & $2.0 \times 10^{7}$ & $3.0 \times 10^{6}$ & 0.34 & 9098
& $1.01 \times 10^9$ \\
Halo C & $1.4 \times 10^7$  & $1.2 \times 10^7$ & $1.6 \times 10^6$ & 0.30 & 7402 
& $1.18 \times 10^9$ \\
\hline 
\hline

\end{tabular}
\parbox[t]{0.9\textwidth}{\textit{Notes:} The above table contains the simulation name$^a$, 
  the total mass$^{b}$ (gas \& dark matter) at the virial radius$^4$ [$M_{\odot}$], 
  the enclosed mass in dark matter$^{c}$ at the virial radius [$M_{\odot}$], the enclosed mass 
  in baryons$^{d}$ at the virial radius [$M_{\odot}$], the virial radius$^e$ [kpc], the virial 
  temperature$^{f}$ [K], and the maximum gas number density$^{g}$ in the halo [$\rm{cm^{-3}}$]. 
  All units are physical, unless explicitly stated otherwise.}

\label{Table:Outer}

\end{table*}


\subsection{Particle Splitting}
In order to increase the resolution of the dark matter particles still further we employ the
particle splitting techniques of \cite{Kitsionas_2002}, several other authors have previously 
adopted the same technique \citep{Yoshida_2006, Greif_2011, Kim_2011, Hirano_2014}. We split 
particles at a redshift of $z = 40$ so as to minimise the effects of any noise introduced by 
the splitting technique. In particle splitting, the distribution of matter within a region 
assumes that the region has a uniform mass distribution and so no new perturbation modes are added. 
Splitting the particles at a redshift of 40 ensures that we approximate this case and 
furthermore we test for convergence as described below. \\
\indent Particles are either split once or twice. Each splitting reduces the particle 
mass by a factor of 13 producing 13 new child particles in the process and so also increasing the 
computational demands of the simulation. We follow \cite{Kitsionas_2002} in choosing to split 
each parent particle into 13 child particles, the choice of 13 is somewhat arbitrary but attempts to 
take into account two competing arguments - 1) there should not be too large a difference
between the child mass and the parent mass and 2) the collective density distribution of the 
family of children should approximate the spherically symmetric density distribution of the parent.\\
\indent We test for convergence using both one and two iterations of splitting. We split the 
particles in a subregion surrounding the point of maximum density. The location of the 
point of maximum density is found from previous simulations where particle splitting was not used. 
This allows for the increased mass resolution to be targeted in the region surrounding the collapse. 
We select a region of $43.75$ h$^{-1}$ kpc (comoving) within which to do the first iteration 
of particle splitting. The second splitting, if performed, is done within a region half this size - 
$21.88$ h$^{-1}$ kpc (comoving). This allows for maximum dark matter resolution alongside 
maximum grid resolution. \\
\indent In order to investigate any systematic effect on the dark matter particle properties we 
initially conducted dark matter only simulations where we split the particles at a redshift of 
$z = 40$ and subsequently looked at the highest density region approximately 80 Myrs later at a 
redshift of $z = 23$. In the runs with baryons the collapse occurs at a redshift of 
approximately $z = 23$ or earlier and so examining the forming dark matter halo at a 
redshift of $z = 23$ is appropriate. \\ 
\indent In Figure \ref{Splitting} we show the dark matter density profiles and enclosed mass 
profiles for Halo A at a redshift of $z = 23$. The profiles are all centred at the point of 
maximum density as found in the simulation where particle splitting was not conducted (i.e. 
the simulation with a dark matter particle mass of M$_{\rm{DM}}$ = 122.7 \msolarc). In the left hand
panel the density profile shows clearly 
that the differences between the run with no particle splitting (blue line) and the runs with 
particle splitting (cyan and magenta lines) are negligible which gives us confidence that the 
dark matter splitting introduces little or no additional perturbing effects to the dark matter 
component. This should not be surprising since the splitting occurs early in the halo formation 
process at a time when the non-linear effects of the gravitational collapse are still negligible. 
To further emphasise the point we show the difference between the enclosed mass profiles
for the same outputs. We normalise against the the highest resolution simulation (i.e. the 
simulation with a dark matter particle mass of M$_{\rm{DM}}$ = 0.7 \msolarc). The differences are 
negligible at large radii (R $\gtrsim 100$ pc) and increase to only a factor of $\lesssim 0.3$ 
at small scales. \\
\indent Figure \ref{Splitting_Visualisation} shows a density 
projection of the dark matter density along the x-axis for this halo (Halo A) for each case. 
The left hand panel has a minimum dark matter particle mass of $M_{\rm{DM}} = 122.7$ \msolarc, the
middle panel has a minimum dark matter mass of $M_{\rm{DM}} = 9.5$ \msolar and the right hand panel 
has a minimum dark matter mass of $M_{\rm{DM}} = 0.7$ \msolarc. The halo becomes smoother as the dark 
matter resolution is increased but otherwise remains visually the same. The spatial scale in 
each case is given in the left panel and corresponds to 250 pc physical.

\subsection{Particle Smoothing} \label{Sec:ParticleSmoothing}
In regions of high density (resolution) the cell size can become very small and the particles 
can begin to experience spurious two body effects. In order to avoid such spurious interactions the 
dark matter particles can be ``smoothed'' onto the grid once the cell size drops below 
a given threshold. In \enzo this is controlled via the \emph{MaximumParticleRefinementLevel} 
parameter. Once the cell size drops below a user defined threshold the dark matter mass is 
interpolated onto the grid with some smoothing length $h$ (which will always be quoted in 
comoving units). While computationally inefficient to do, it is done in only a very 
small region and so its effect on the overall computational time is small. The dark
matter now effectively behaves as a continuous, rather than discrete, distribution in
these very high density regions and the effects of (large) dark matter particles coming into 
close contact is alleviated. In this study we set the threshold value of $h$ to 2.8 \emph{comoving} 
parsecs (equivalent to a refinement level of 12). We run simulations where the smoothing is 
initiated and also where the smoothing is not activated. In the runs without the smoothing the 
smoothing length is effectively the minimum cell size (i.e. the cell size at maximum refinement 
level) which in the case of our simulations is 0.04 \emph{comoving} parsecs at level 18.
\begin{figure*}
  \centering 
  \begin{minipage}{175mm}      \begin{center}
      \centerline{
        \includegraphics[width=18cm]{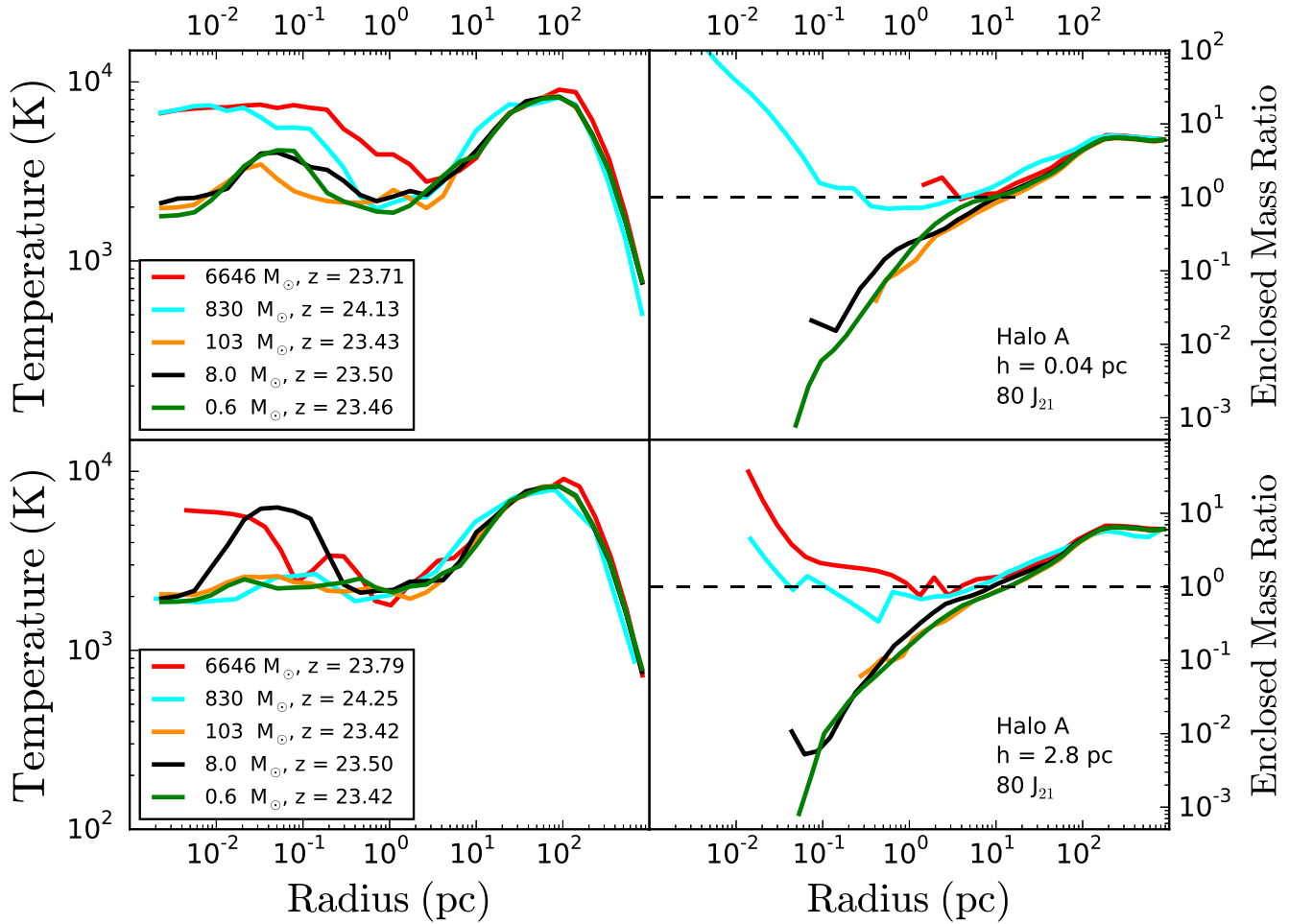}}
        \caption[]
        {\label{HaloA:MultiPlot_J80}
          \textit{Halo A}: In the top two panels no smoothing of the dark matter particles is 
          activated and so the smoothing length is the same as the minimum cell 
          size - 0.04 pc (comoving). In the bottom two panels the smoothing is activated 
          at a refinement level of 12 (2.8 pc comoving). The bottom right panel shows the 
          enclosed dark matter mass divided by the enclosed baryon mass - 
          $\rm{M_{\rm{DM}}/M_{Baryon}}$. See text for further details on the plot characteristics. 
          }
      \end{center} \end{minipage}
  \end{figure*}

\subsection{External Radiation Fields}
As noted above we concentrate our study on external radiation fields with values of 
80 J$_{21}$ and 500 J$_{21}$ both of which may be relevant for studying the direct collapse of 
super-massive black holes at high redshift \citep{Shlosman_1989, Begelman_2006, Dijkstra_2008, 
Regan_2009, Regan_2009b, Shang_2010, Johnson_2013b, Latif_2013c, Latif_2014a, Latif_2014b, 
Agarwal_2014b, Regan_2014a, Regan_2014b}. A value of 80 J$_{21}$ sets the field to where \molH 
formation is inhibited but not completely destroyed, the collapse of the halo therefore takes 
place in an environment where the collapse is not isothermal. A value of 500 J$_{21}$ on the 
other hand results in an isothermal collapse. By examining both scenarios the effects of 
different dark matter resolutions on the direct collapse model can be fully assessed. \\
\indent We further assess the impact of smaller radiation fields on the collapse. We use field 
strengths of both 1 J$_{21}$ and 10 J$_{21}$ and we also test the case where no background exists 
(i.e. 0 J$_{21}$). All of these fields may be relevant for the case where metal-free 
stars form \citep[e.g.][]{Bromm_1999, Abel_2002, Bromm_2002}. A comparison of the affect of 
different field strengths is discussed in \S \ref{Sec:Fext}. \\
\indent We do not examine the affect of changing the \molH three body formation rates in this 
work and instead use the rates of \cite{Martin_1996} throughout. This topic has been 
examined comprehensively by both \cite{Turk_2011} and \cite{Bovino_2013}. The \cite{Martin_1996}
rates are found to provide the most consistent fit from the work of \cite{Turk_2011}, are density
dependent and follow closely the rates of \cite{Forrey_2013} which are advocated by 
\cite{Bovino_2013}.
\footnotetext[4]{The virial mass is 
    defined as 200 times the mean density of the Universe in this case.}
\section{Simulation Details} \label{Sec:Simulations}
The primary goal of this study is to evaluate the impact, if any, of the dark matter resolution 
on the results of the collapse of the gas. The general halo characteristics of Halo A, Halo B and
Halo C are given in Table \ref{Table:Outer}. The study focuses on Halo A for the most part 
with the characteristics of Halo B and Halo C only discussed in \S \ref{Sec:CompareBC}. \\
\indent Each (re)simulation of Halo A differs in either the minimum dark 
matter particle size or the minimum smoothing length of the dark matter particle. Furthermore, 
the external radiation field that a simulation is exposed to is varied between 0 J$_{21}$ and 
500 J$_{21}$. The simulations are automatically halted once the maximum refinement level is reached. 
Therefore each simulated halo, which is exposed to the same external radiation field, also has a 
very similar maximum density and comparisons are made at this maximum refinement level output. 
Each simulation is therefore compared at a very similar stage in its evolution. \\
\indent The smoothing lengths for each run are set either at the maximum refinement level 
which corresponds to 0.04 (refinement level 18) comoving pc or they are smoothed at a refinement 
level of 12 which corresponds to 2.80 comoving pc (see \S \ref{Sec:ParticleSmoothing} for more 
details on setting the particle smoothing lengths). Each parameter used in each simulation is 
clearly marked in the left hand plot window of each figure. To promote clarity we refer to a 
simulation with a dark matter particle resolution of 0.6 \msolar as S06, a simulation with a 
dark matter particle resolution of 8.0 \msolar as S8, etc.

\section{Results} \label{Sec:Results}

\begin{figure*}
  \centering 
  \begin{minipage}{175mm}      \begin{center}
      \centerline{
        \includegraphics[width=18cm]{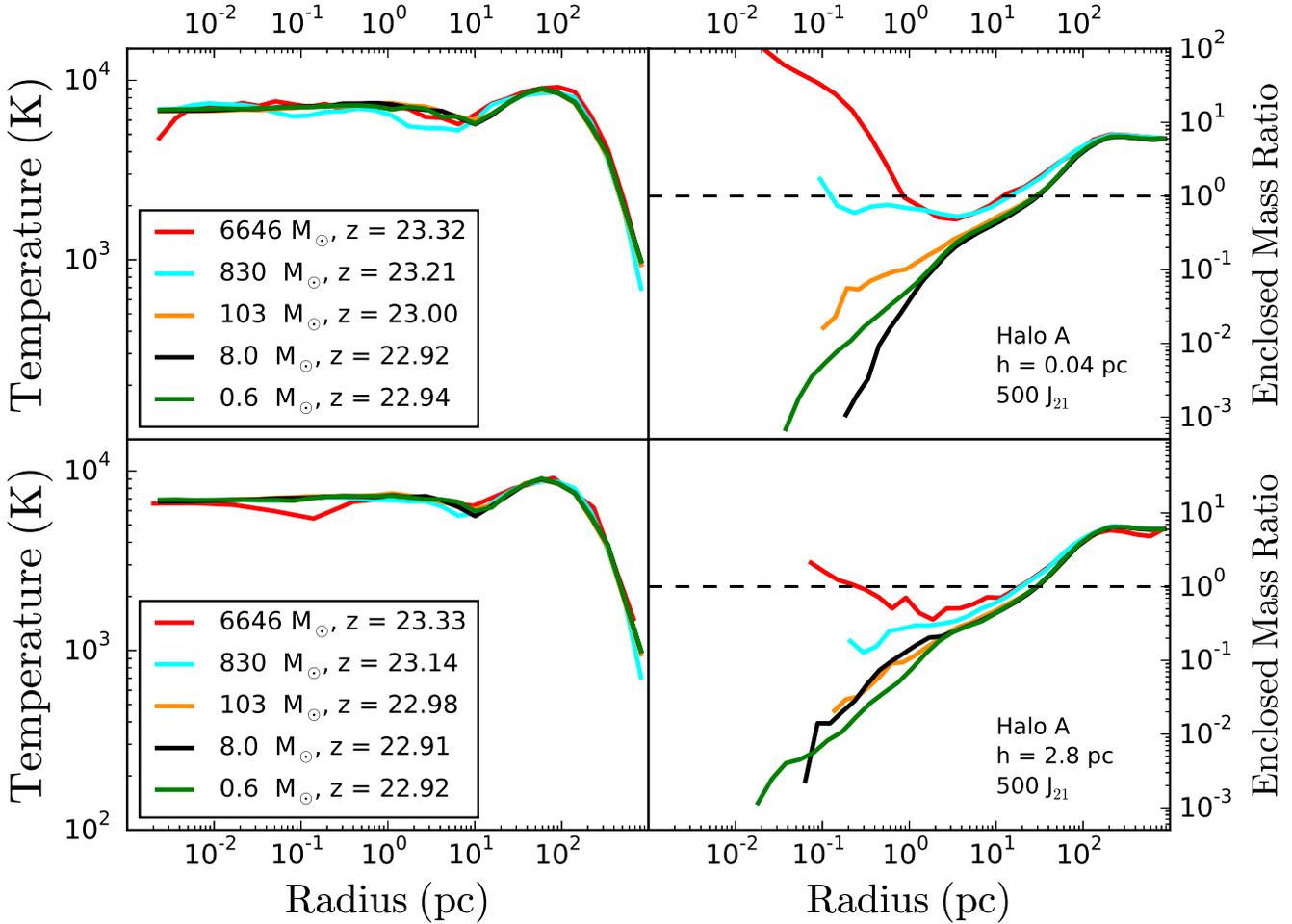}}
      \caption[]
      {\label{HaloA:MultiPlot_J500}
        \textit{Halo A}: Same as Figure \ref{HaloA:MultiPlot_J80} except 
        the external radiation field is set to 500 J$_{21}$. 
      }
    \end{center} \end{minipage}
\end{figure*}

\subsection{Halo A - Varying the Mass Resolution and Smoothing Lengths} \label{Sec:Smoothing}
We begin by examining the temperature profile for the collapsed haloes for the different dark matter 
mass resolutions and smoothing lengths. In the following cases the field strength is set to either 
80 J$_{21}$ or 500 J$_{21}$, we will examine the case of lower background radiation intensities 
in \S \ref{Sec:Fext}. 

\indent In Figure \ref{HaloA:MultiPlot_J80} we show the results for Halo A when the external 
field is fixed at 80 J$_{21}$. The top row in the panel shows simulations where no smoothing of 
the dark matter particles occurs. The bottom panel shows simulations where the smoothing is 
activated at a refinement level of 12 (2.8 comoving parsecs). Each simulation was run with 
different minimum dark matter particle resolution ranging from M$_{\rm{DM}} = 6646$ \msolar to 
M$_{\rm{DM}} = 0.6$ \msolarc. Looking at the left hand column where we have plotted the temperature 
profiles, centred on the densest point in the simulation, we see that down to approximately 1 parsec
from the centre the smoothed and non-smoothed simulations agree very well. Within the core of the 
halo differences begin to appear. In the non-smoothed simulations (top left panel) the two lowest
resolution simulations show a hotter core compared to the higher resolution simulations (i.e. 
those with a particle resolution M$_{\rm{DM}} \le 103$ \msolarc). In the smoothed runs the lowest 
resolution simulation has a temperature in the core of T$_{\rm{core}} \sim 6000$ K while all the other 
simulations converge to a core temperature T$_{\rm{core}} \sim 2000$ K - note also that S8 shows 
some significant heating within the core at a radius of approximately 0.05 pc. It should 
also be noted here  that we do not include the effects of radiation self-shielding 
in our simulations which may be important in the core of the halo.\\
\indent The right hand column yields a partial explanation for this. Considering first the 
non-smoothed runs in the top right panel - in both S830 and S6646 the core is dark 
matter dominated. The right hand panel plots the ratio of the enclosed dark matter mass to the 
enclosed gas mass. Values greater than 1.0 indicate regions where the dark matter dominates 
compared to the gas. In simulations S830 and S6646 the region inside approximately 1 parsec is 
dark matter dominated (in S6646 a single dark matter particle sits approximately 1 parsec from 
the maximum gas density but its mass dominates the region). The results from S830 and S6646 are 
therefore likely to be spurious as the dark matter resolution in each case is insufficient and 
the dark matter mass ends up dominating the central potential. The heating of the core, seen 
in the left hand panel, should therefore be treated with caution due to the presence of perturbing 
forces from these large dark matter particles. \\
\indent For the case of the smoothed simulations we again plot the dark matter to gas enclosed 
mass ratios
in the bottom right panel. In this case we again see that S6646 is dark matter dominated and that 
S830 is only marginally gas dominated. The higher resolution simulations (S103, S8 and S06) on the 
other hand are all strongly gas dominated. Again in this case S6646 and S830 should be treated 
with caution (even though S830 shows converged results). \\
\indent Regarding the temperature dichotomy within the core seen in the left hand column, the result 
is not as bad as it first appears and reflects the time at which we choose to examine the core. 
Excluding S830 and S6646 we see that S06, S8 and S103 all agree quite well (with the possible 
exception of the smoothed S8 simulation which shows a bump at approximately 0.05 pc). However, 
allowing the simulations to run at the maximum refinement level for a short time longer results 
in the core heating up in \emph{each case}. This results from the collisional dissociation of 
\molH which pushes the core onto the atomic cooling track once again - this is shown in the left 
panel of Figure \ref{HaloA:Extra} and discussed in more detail in Appendix \ref{appendixA} 
(as it lies somewhat outside the scope of the current work). We further caution the reader that 
running at the maximum refinement level means we are artifically preventing collapse which may 
induce unphysical results and that further research needs to be done to examine this scenario 
in more detail. \\
\indent Strong convergence between  S06, S8 and S103 is therefore achieved, when the external 
field is set to 80 J$_{21}$, but the study indicates that the lower resolution simulations 
dissociate the \molH faster compared to the higher resolution simulations, driven by spurious 
dark matter effects, which impact the hydrodynamics. \\
\indent In Figure \ref{HaloA:MultiPlot_J500} we show the results when the background radiation is 
set to 500 J$_{21}$ with all other parameters remaining unchanged. In this case the external 
radiation field is strong enough to reduce the \molH fraction by several orders of magnitude within 
the halo. The left hand column of Figure \ref{HaloA:MultiPlot_J500} shows the temperature profiles 
which show highly converged results for both the non-smoothed (top panel) and smoothed (bottom 
panel) for even the lowest dark matter resolutions. The right hand column shows the dark matter to 
gas enclosed mass ratio. The S6646 simulations are dark matter dominated in both cases (and yet 
the temperature profiles remain converged) but the S830 simulation is more gas dominated 
in the smoothed simulations compared to the non-smoothed case. The higher resolution runs also 
show stronger convergence when smoothed. 

\begin{figure*}
  \centering 
  \begin{minipage}{175mm}      \begin{center}
      \centerline{
        \includegraphics[width=18cm]{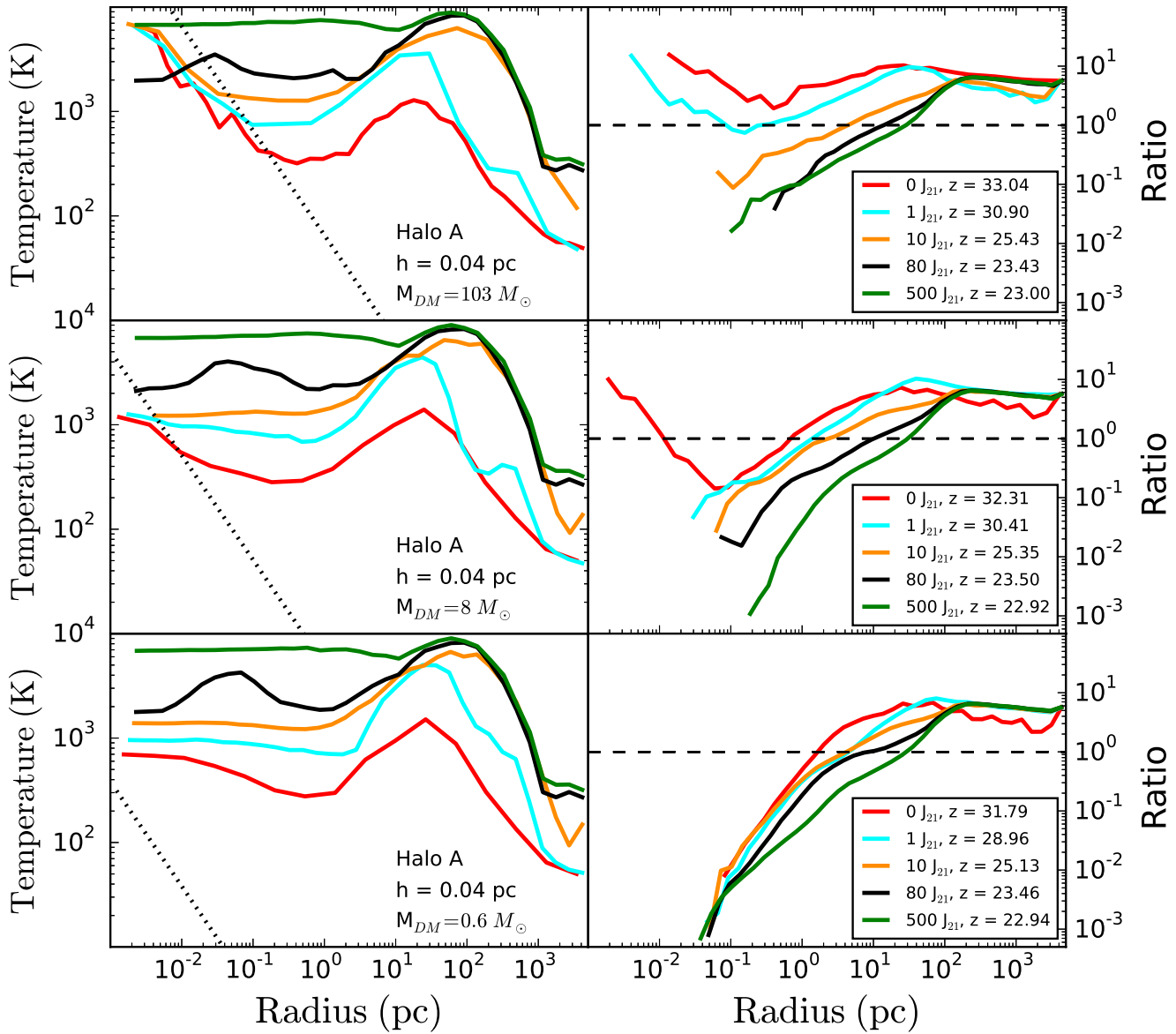}}
      \caption[]
      {\label{HaloA:MultiPlot_SixPanel}
        \textit{Halo A}: Each row shows the temperature profile and the ratio of the
        dark matter enclosed mass to the baryon enclosed mass. The y-axis label is simply 
        identified as ``Ratio'' for convenience. The top row shows the case where the particle mass
        is set to M$_{\rm{DM}} = 103$ \msolar and no smoothing is employed. The 
        dark matter mass resolution then varies between each row with every other parameter 
        remaining unchanged. In each individual panel the external field varies between 0 J$_{21}$
        and 500 J$_{21}$. The dotted line in the left hand panels is the ``discretisation 
        temperature'' as defined in equation \ref{Tdisc}. At a given scale it shows the effect
        a dark matter particle of a given mass can have on the baryon temperature.
      }
  \end{center} \end{minipage}
\end{figure*}

\subsection{Halo A - Comparing Smoothed and non-Smoothed Simulations}
\indent The effect of smoothing the particles onto the grid is most noticeable in the low resolution 
simulations (i.e. S6646 \& S803) while the effects become less pronounced
once the dark matter particle mass drops or equivalently once the discretisation velocity, 
V$_{\rm{disc}}$, becomes less than the local sound speed (equal to a few km/s at these temperatures). 
We define V$_{\rm{disc}}$ roughly as the velocity that a dark matter particle can impart onto its 
surroundings
\begin{equation}
\rm{V_{disc} = \sqrt{G M_{\rm{DM}} \over R_s}}
\end{equation}
where G is the usual gravitational constant, M$_{\rm{DM}}$ is the mass of a single dark matter 
particle and R$_s$ is the smoothing length.  This relationship can also be recast in terms of 
a discretisation temperature resulting in the equation
\begin{equation}
\rm{T_{disc} = \sqrt{\mu m_H G M_{\rm{DM}} \over R_s \ k_b}}
\end{equation}
where $\mu$ is the mean molecular weight, set to be 1.22 in this study, and $\rm{k_b}$ is the 
Boltzmann constant. This equation can be further expanded by replacing R$_S$ with max(R, R$_S$) 
where R is any radius becoming
 \begin{equation} \label{Tdisc}
\rm{T_{disc}(R) = \sqrt{\mu m_H G M_{\rm{DM}} \over max(R, R_S) \ kb}}
\end{equation}
Equation \ref{Tdisc} can then be used to determine the radius at which, for a given dark matter 
particle mass, the dark matter (under-)resolution can begin to influence the baryon temperature. 
We will return to this point in \S \ref{Sec:Fext}.\\
\indent If we compare the non-smoothed simulations to the smoothed ones when the radiation field is 
set to 80 J$_{21}$ (i.e. upper and lower rows of Figure \ref{HaloA:MultiPlot_J80})
we see that the temperature profiles in the left hand panels both show a significant scatter.  
The mass ratios are qualitatively similar (right hand panels) in both the smoothed and 
non-smoothed runs with the higher resolution (S103, S8 and S06) well converged but with S6646 and 
S830 showing significant scatter between the smoothed and non-smoothed runs. For example, 
the S830 run in the smoothed simulation is marginally gas dominated while in the non-smoothed 
case S830 is strongly dark matter dominated. The overall effect of smoothing when the radiation 
field is set to 80 J$_{21}$ is therefore marginal, the higher resolution simulations are already 
well converged and so smoothing only improves the convergence slightly. The lower resolution 
runs are dark matter dominated and smoothing cannot prevent this although smoothing may somewhat 
lessen the adverse effects. \\
\indent In comparing the cases where the external radiation field is stronger with J set to 
500 J$_{21}$ the effects are only slightly more encouraging. In the 500 J$_{21}$ case the gas is 
warmer and hence the sound speed is correspondingly higher meaning that the discreteness effects, 
quantified by V$_{\rm{disc}}$ are somewhat lessened. This has the effect that for the temperature 
profiles the results appear completely converged as the effects of the radiation overwhelm any 
spurious dark matter effects. Nonetheless in both the 80 J$_{21}$ case and the 500 J$_{21}$ 
case the S6646 simulation is strongly dark matter dominated and smoothing the dark matter onto 
the grid at this stage only alleviates slightly this particular ill effect. However, in the case 
of S830 we observe that while in the non-smoothed case the core is marginally gas dominated it is 
significantly gas dominated in the smoothed case - although the ratio remains above 0.1. 
Finally similar to the 80 J$_{21}$ case the simulations with a particle mass 
M$_{\rm{DM}} \lesssim 103 $ \msolar are relatively well converged in both the smoothed and 
non-smoothed cases although the smoothed case does show less deviation at the smallest scales. \\
\indent Therefore, the result is that smoothing tends to improve convergence only in isolated cases 
or when the dark matter resolution is already high. However, smoothing poorly 
resolved dark matter particles onto the grid will likely have no positive effect and it is therefore
more important to adequately resolve a given region with a sufficiently high dark matter mass 
resolution \emph{and} then to smooth the particles onto the grid to remove any lingering discrete
effects from the dark matter particles.

\begin{figure*}
  \centering 
  \begin{minipage}{175mm}      \begin{center}
      \centerline{
        \includegraphics[width=18cm]{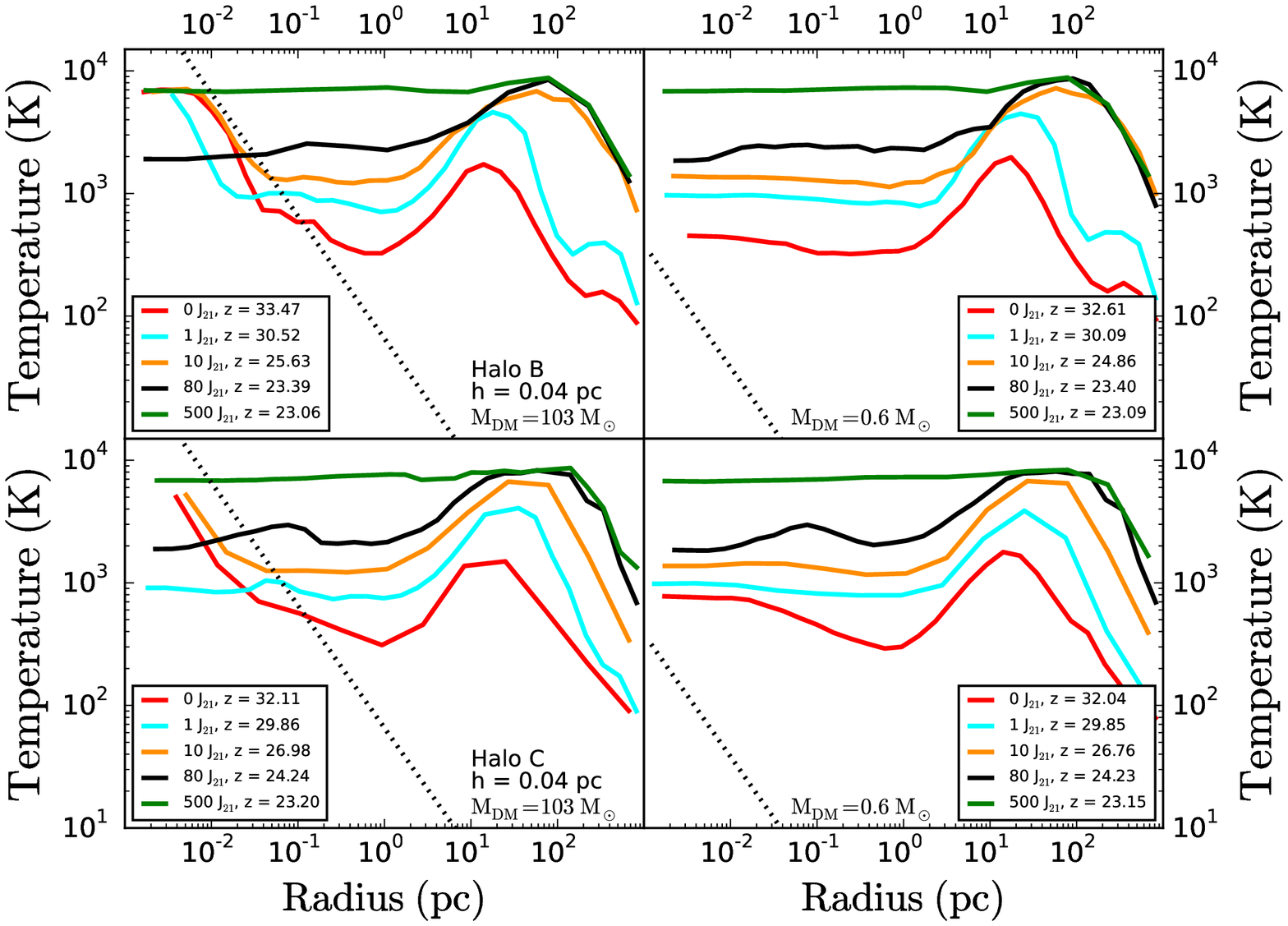}}
      \caption[]
      {\label{CompareBC}
        \textit{Halo B \& Halo C}: The temperature profile for haloes B \& C are shown for 
        two different dark matter particle resolutions. Halo B is shown in the top row while halo 
        C is displayed in the bottom row. The particle resolution is set at M$_{\rm{DM}}$ = 103 
        \msolar in the left hand column while it is set to  M$_{\rm{DM}}$ = 0.6 \msolar in the right 
        hand column. The dotted line in each panel is our analytical estimate of the 
        discretisation temperature, $\rm{T_{disc}(R)}$.
      }
  \end{center} \end{minipage}
\end{figure*}

\begin{figure*}
  \centering 
  \begin{minipage}{175mm}      \begin{center}
      \centerline{
        \includegraphics[width=9cm]{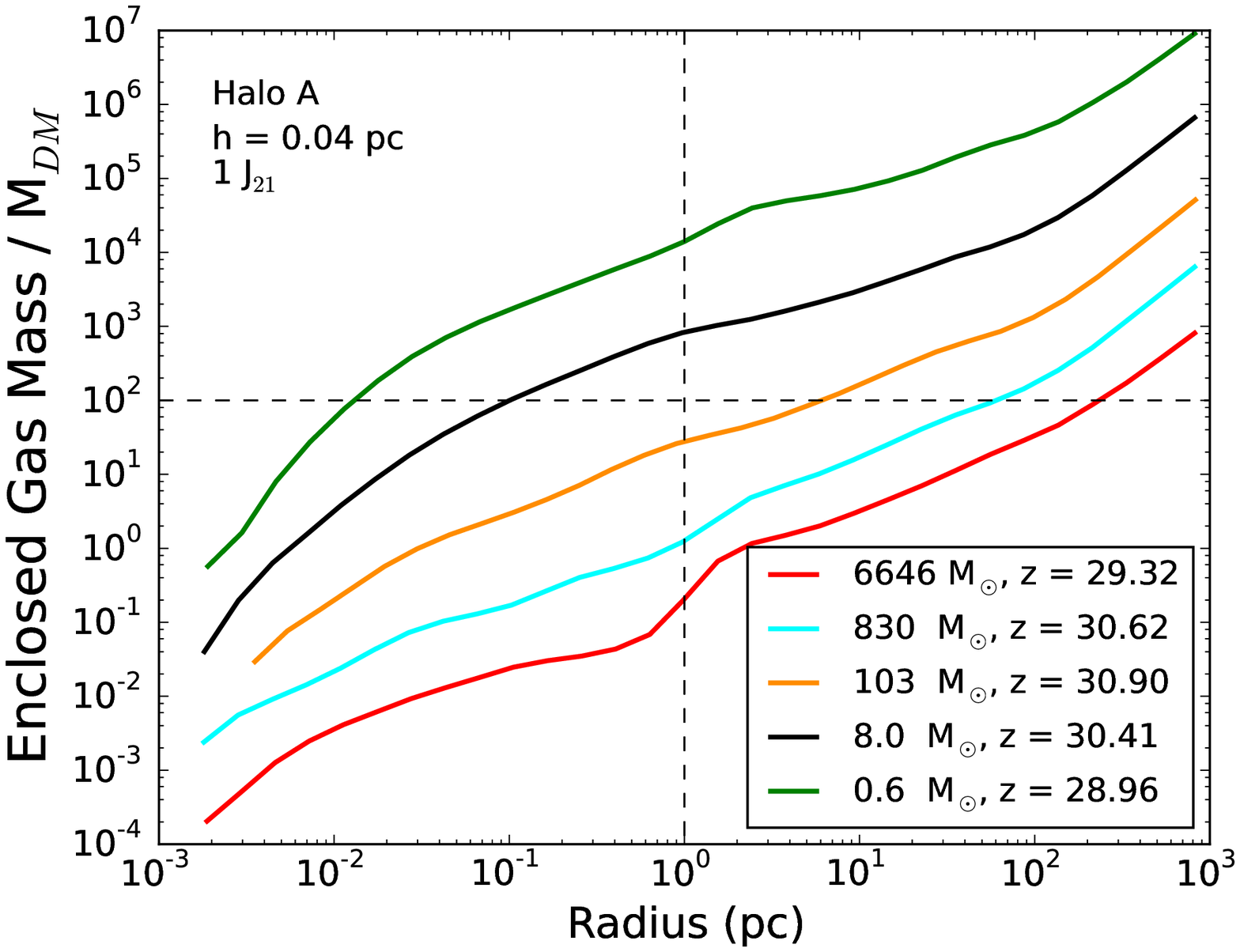}
        \includegraphics[width=9cm]{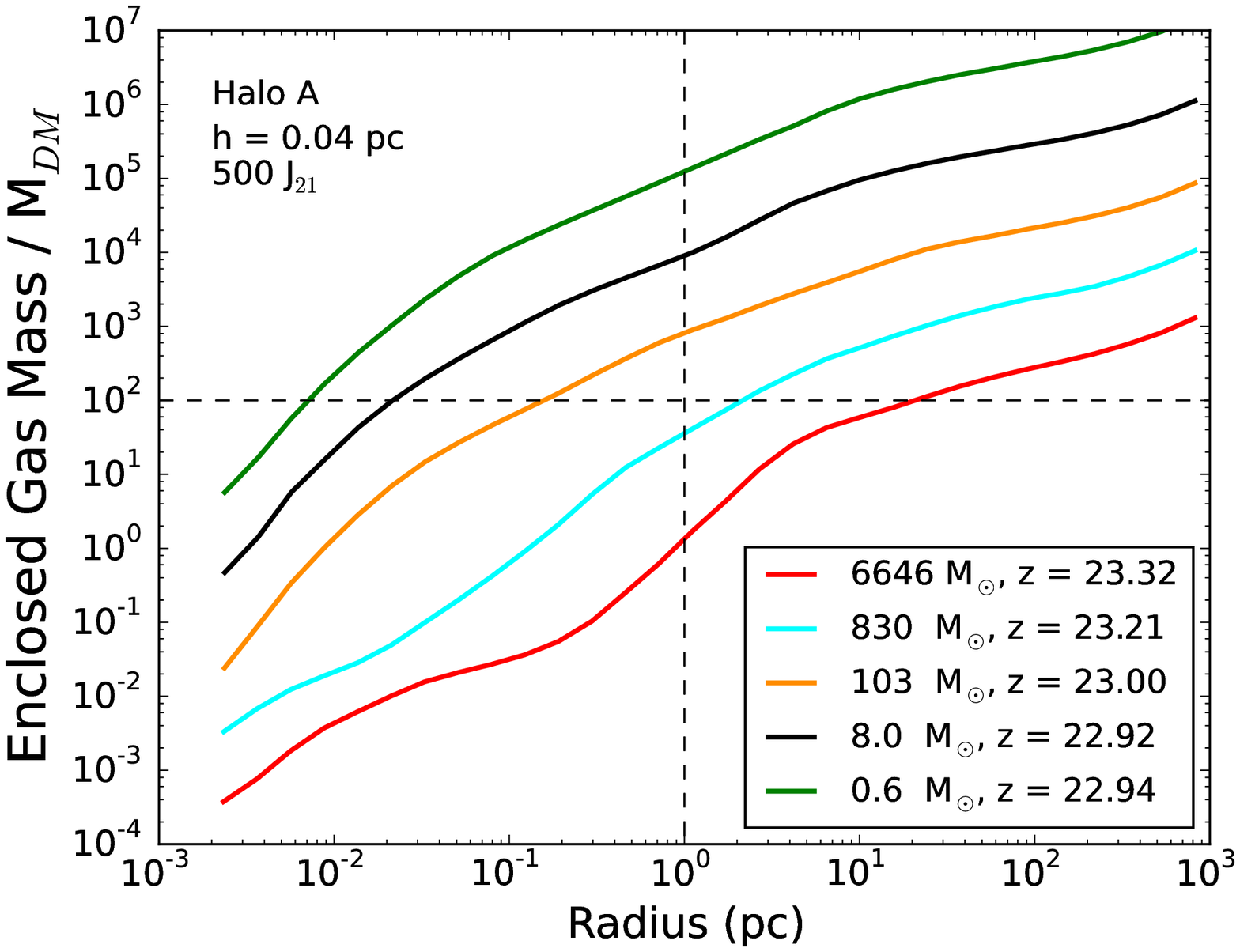}}
      \caption[]
      {\label{HaloA:MassScaling}
        \textit{Halo A}: Radial profiles of the enclosed baryonic mass divided by the 
        mass of the (minimum) dark matter particle mass, M$_{\rm{DM}}$, used in each simulation. 
        In order to satisfy our resolution requirement the ratio of the 
        enclosed baryonic mass to the dark matter particle mass should exceed 
        100.0 at a radius of 1 parsec. In the left hand panel the external radiation 
        field is set to 1 J$_{21}$ while in the right hand panel the radius is set
        to 500 J$_{21}$. The dashed lines marks the radius at 1 pc (vertical dashed line)
        and the ratio value of 100 (horizontal dashed line). 
      }
  \end{center} \end{minipage}
\end{figure*}

\subsection{Halo A - Changing the External Field Amplitude} \label{Sec:Fext}
So far we have selected only external fields of 80 J$_{21}$ or 500 J$_{21}$. These fields 
are relevant for investigating the case of a collapsing halo exposed to a 
nearby strong UV source as may be required in the direct collapse formation model
\citep[e.g.][]{Dijkstra_2008, Visbal_2014b}. We now turn our 
attention to investigating the effects of other external field amplitudes which may be expected at
this redshift. In particular, we choose three further field amplitudes of 0 J$_{21}$ (i.e. no 
background radiation), 1 J$_{21}$ and 10 J$_{21}$. To test for any discreteness effects from
under-resolved dark matter particles we do not smooth the particles onto the grid so the 
smoothing length is set to the cell size of the finest grid - 0.04 comoving parsecs in each case. \\
\indent In Figure \ref{HaloA:MultiPlot_SixPanel} we have plotted the radial profiles of the 
temperature and the dark matter mass to gas ratio with the mass resolution 
increasing between each panel as indicated in the right hand column of the figure. \\
\indent Examining first the top panel where the dark matter particle resolution is set to 
M$_{\rm{DM}}$ = 103 \msolar we see that the simulations with lower field amplitudes (0 J$_{21}$, 
1 J$_{21}$ and 10 J$_{21}$) all show a strong temperature increase towards the core of the halo.  
As the mass resolution is increased (middle and bottom rows) this phenomenon disappears and the 
temperature profiles converge to temperatures close to $T \sim 1000$ K as would be expected for the 
formation of metal-free stars at this redshift \citep{Abel_2002, Yoshida_2006, Yoshida_2008, 
Bromm_2009, Smith_2009, Turk_2009, Stacy_2010, Clark_2011, Stacy_2014}. \\
\indent In the lowest mass resolution simulations this spurious behaviour can be explained by 
examining the dark matter to gas ratio in the right hand column. In the lowest mass resolution 
simulations (M$_{\rm{DM}}$ = 103 \msolarc, top right panel) the simulations with external radiation 
amplitudes of 0 J$_{21}$, 1 J$_{21}$ and 10 J$_{21}$ are all dark matter dominated (or only mildly 
gas dominated) within the core of the halo resulting in spurious heating of the gas due to the 
presence of large dark matter particles (we saw this before in Figure \ref{HaloA:MultiPlot_J80} for 
example). As we increase the dark matter particle resolution we see this behaviour disappear as 
the dark matter to gas ratio drops significantly and the temperature profiles converge to 
$T \sim 1000 $ K as expected. \\
\indent In the temperature panels (left hand panels) we have also overplotted our analytical 
estimation of the discretisation temperature, $\rm{T_{disc}(R)}$, (see equation \ref{Tdisc}). This 
gives the scale at which we expect the dark matter particle mass to start to impact (negatively) on 
the temperature. In the top left panel we see that the large dark matter particle mass used here 
(M$_{\rm{DM}}$ = 103 \msolarc) means that the $\rm{T_{disc}(R)}$ function intersects the 0 J$_{21}$ 
profile very close to where the baryon temperature starts to show a spurious increase. A similar 
effect is seen in the middle left panel where the particle mass is set to (M$_{\rm{DM}}$ = 8 \msolarc).
In the bottom left panel the  $\rm{T_{disc}(R)}$ function does not intersect any of
the temperature profile lines. This indicates that at this particle resolution we would not expect 
any impact from dark matter heating effects at the spatial scales probed in our simulations and 
this is exactly what we see. \\
\indent Taking the field strength of 1 J$_{21}$ as the fiducial case we find that the core ($<$ 1 pc)
of the simulation was resolved by only 36 dark matter particles (with a V$_{\rm{disc}} \sim 18$ km/s 
and T$ _{\rm{disc}} \sim 47000$ K) when the mass resolution was set to M$_{\rm{DM}}$ = 103 \msolarc. 
When the resolution was increased to M$_{\rm{DM}}$ = 8 \msolar the number of dark matter particles 
in the core increased to 473 particles (V$_{\rm{disc}} \sim 5$ km/s and T$_{\rm{disc}} \sim 3600$ K) and 
increasing the resolution to M$_{\rm{DM}}$ = 0.6 \msolar resulted in the core being resolved by 
3072 particles (V$_{\rm{disc}} \sim 1$ km/s and T$_{\rm{disc}} \sim 300$ K). \\
\indent The smaller haloes, those with a gas core less than $\sim 10^4$ \msolarc, which collapse 
when the external field amplitude is relatively small require a much higher dark matter resolution. 
In this case we find that the dark matter particle size must be less than $\sim 8 $ \msolar 
(although even at M$_{\rm{DM}} \sim 8 $ \msolar we see some spurious heating of the core for the case 
of J = 0 J$_{21}$) and ideally less than $\sim 1 $ \msolar in order to ensure that the core of the 
halo is dominated by baryons and not by spuriously large dark matter particles. 

\subsection{Comparing Halo B and Halo C} \label{Sec:CompareBC}
We now briefly examine the hydrodynamic properties of Halo B and Halo C. In Figure \ref{CompareBC}
we have plotted the temperature profiles of Halo B and Halo C. Halo B and Halo C, similar to Halo A, 
represent high sigma peaks in the density field. As a result they contain a halo which collapses 
early even when exposed to a strong external radiation field i.e. the collapse redshift when the 
external field is set to 500 J$_{21}$ is z$_{\rm{col}} \sim 23.1$ for Halo B and z$_{\rm{col}} \sim 23.2$ 
for Halo C. The top row of Figure \ref{CompareBC} contains plots from Halo B and the bottom row is 
for Halo C. The mass resolution is set at M$_{\rm{DM}}$ = 103 \msolar in the 
left hand column and M$_{\rm{DM}}$ = 0.6 \msolar in the right hand column as indicated. No smoothing 
of the dark matter particles is employed. The external field is varied in each panel to show the 
effect that the external radiation field and the dark matter resolution has on the forming halo in 
each case. Finally, we also overplot $\rm{T_{disc}(R)}$ (equation \ref{Tdisc}) as a 
dotted line in each panel.  \\
\indent Both haloes display very similar properties to Halo A 
(Figure \ref{HaloA:MultiPlot_SixPanel}). For low field strengths (J $\lesssim $ 10 J$_{21}$) and  
M$_{\rm{DM}} = 103 $ \msolar the temperature increases strongly within the core of the halo resulting 
in temperatures close to T $\sim 10^4$ K. This spurious heating in the core of the halo is caused 
by a dark matter particle mass that is too coarse. The large dark matter particle masses induce
perturbing forces on the gas which heats the core. This conclusion is supported by our analytical 
estimate of the heating effect given by $\rm{T_{disc}(R)}$. In left hand panels, where the 
mass resolution is comparatively poor, the temperature profiles for the low J$_{21}$ values begin 
to increase in the core at a radius very close to where the analytical estimate predicts that 
heat may be imparted from dark matter interactions. \\
\indent However, increasing the mass resolution to  M$_{\rm{DM}} = 0.6 $ \msolar resolves this 
spurious heating and in this case the temperatures converge to T $\sim 1000$ K as seen for Halo A in 
Figure \ref{HaloA:MultiPlot_SixPanel}. Again we see this through out analytical estimate of the 
temperature as well. In the right hand panels $\rm{T_{disc}(R)}$ is always below the temperature of the
baryons and hence has no effect on the thermal physics at the scales probed in these simulations.
Similar to the results for Halo A it is clearly demonstrated that 
for small external radiation fields and hence core masses M$_{\rm{core}} \lesssim 10^4$ \msolarc, a 
dark matter particle mass of M$_{\rm{DM}} \sim 1$ \msolar is required. \\
\indent For field strengths with J $\gtrsim $ few $\times \ 10$ J$_{21}$ a very high resolution 
dark matter particle is not required and in these cases a particle mass of $M_{\rm{DM}} \sim 100$ 
\msolar is adequate due to the more massive gas core in these cases.
\begin{figure*}
  \centering 
  \begin{minipage}{175mm}      \begin{center}
      \centerline{
        \includegraphics[width=9cm]{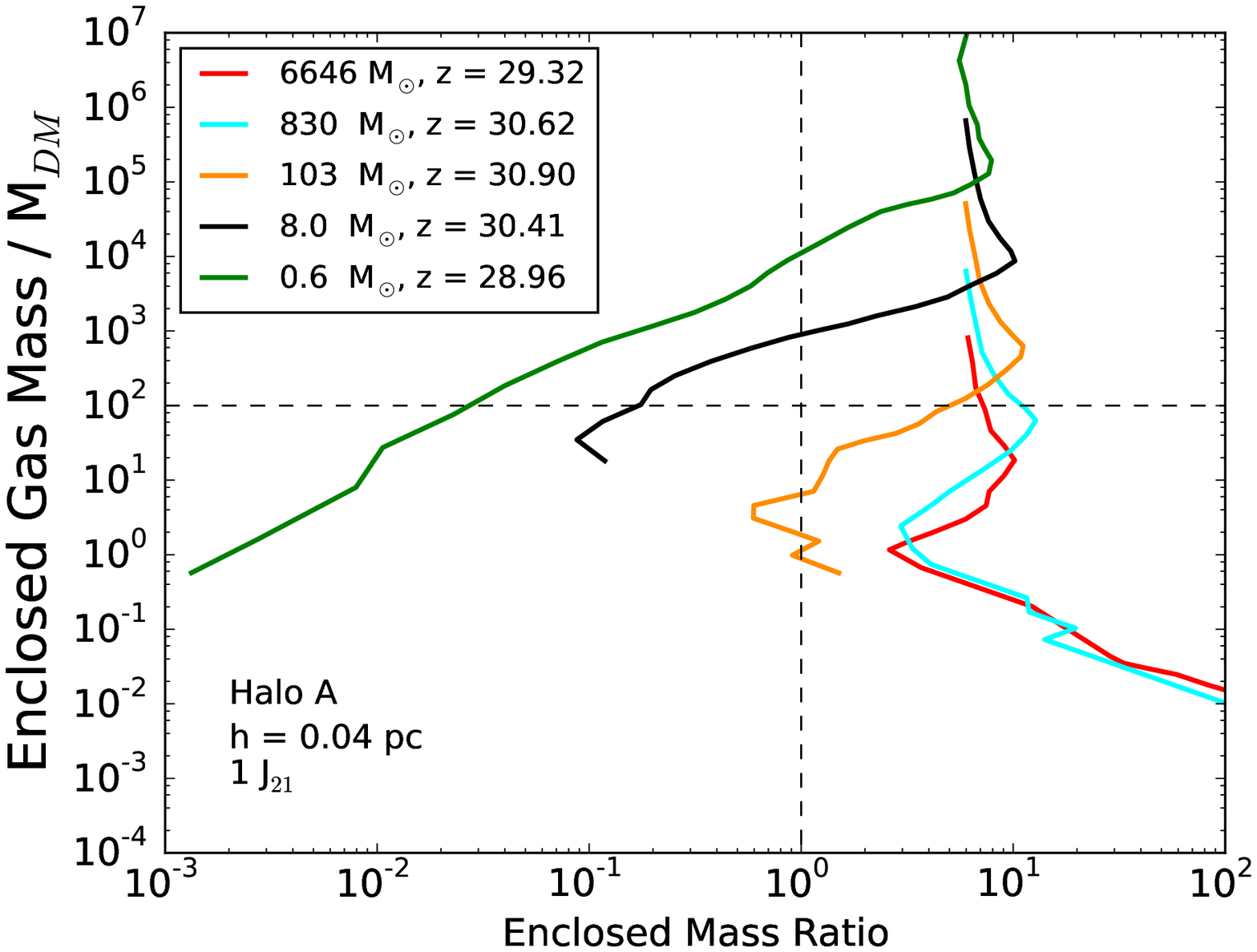}
        \includegraphics[width=9cm]{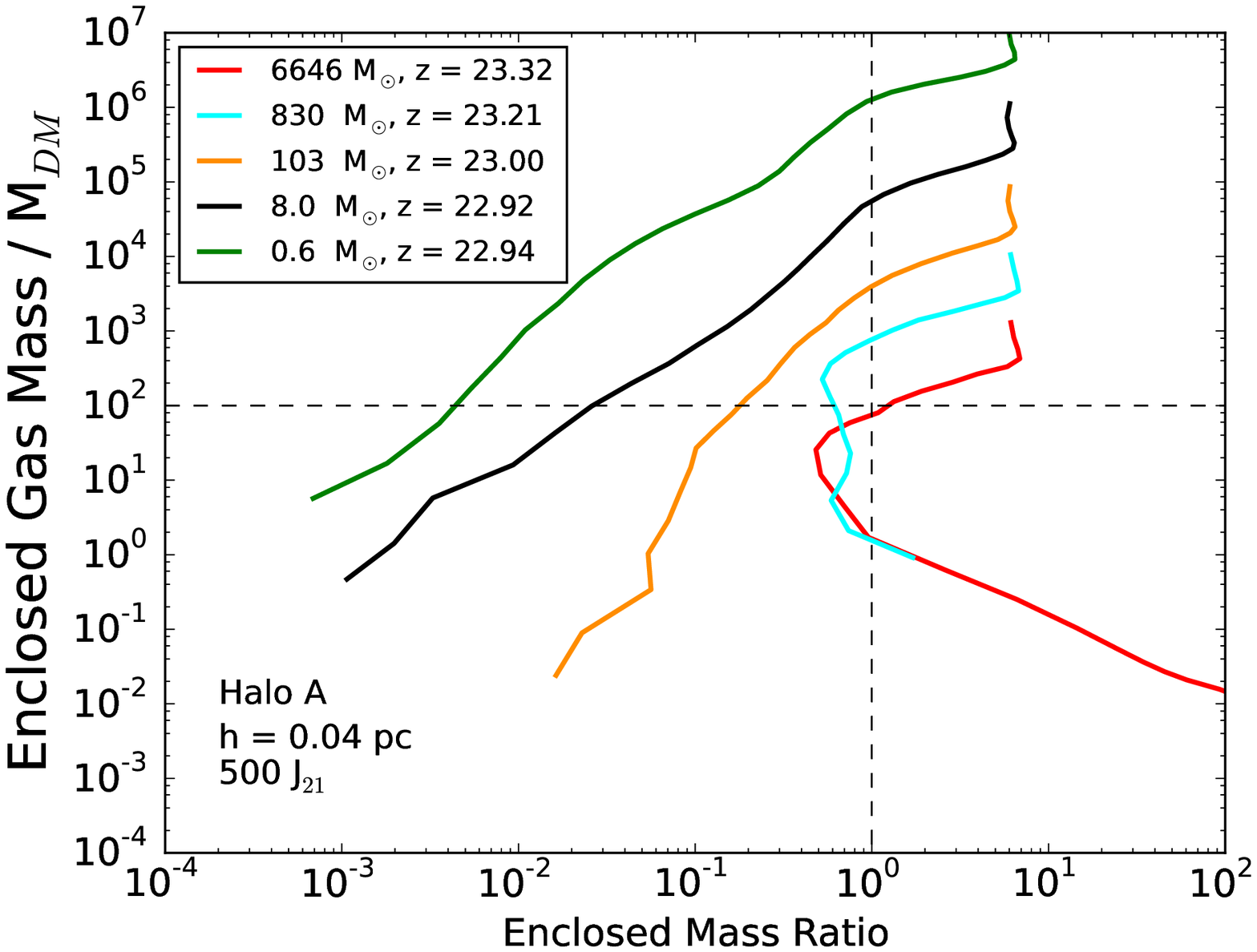}}
      \caption[]
      {\label{HaloA:MassScaling_Mass}
        \textit{Halo A}: Radial profiles of the enclosed baryonic mass divided by the 
        mass of the (minimum) dark matter particle mass, M$_{\rm{DM}}$, used in each simulation.
        The scaling ratio is this time plotted against the enclosed mass ratio (i.e. enclosed 
        dark matter mass divided by enclosed baryon mass).
        The dashed lines marks the point of baryon - dark matter enclosed mass equality 
        (vertical dashed line) and the ratio value of 100 (horizontal dashed line). 
      }
  \end{center} \end{minipage}
\end{figure*}

\section{Choosing the correct dark matter particle size} \label{Sec:Choosing}
Using a suite of simulations we attempted to constrain the dark matter 
resolution required to achieve convergence in the baryonic properties of haloes collapsing
at high redshift. The particle mass required depends only very weakly on the halo (we see no 
differences between the three haloes simulated) but depends strongly on the enclosed 
baryonic mass within the core of the halo (here defined as the radius at which the 
baryonic enclosed mass exceeds the dark matter enclosed mass - typically occuring between 1 parsec 
and 10 parsecs in our simulations). We used different external radiation fields which has the 
effect of varying the mass of the core that forms and the cooling processes available in the 
collapsing halo. \\
\indent In Figure \ref{HaloA:MassScaling} we have plotted the enclosed baryon mass scaled by the 
minimum dark matter particle mass used in each simulation against the radius. We show the results 
for two representative
cases;  J = 1 J$_{21}$ in the left hand panel and J = 500 J$_{21}$ in the right hand panel. From 
our previous results we know that when the external field is set to 1 J$_{21}$ then a
dark matter resolution of M$_{\rm{DM}} < 103.0$ \msolar is required. On the other hand  when the 
field strength is set to 500 J$_{21}$ then a dark matter resolution of M$_{\rm{DM}} < 830.0$ \msolar 
is sufficient. \\
\indent In the case where J = 1 J$_{21}$ we see that the required mass resolutions (S8 and S06) 
have a ratio of the enclosed mass within the core, M$_{\rm{core}}$, to the dark matter particle 
mass, M$_{\rm{DM}}$, of $> 100$ at the core radius i.e.  M$_{\rm{core}}$/M$_{\rm{DM}} >$ 100.0. 
Similarly, in the case where J = 500 J$_{21}$ we see that three dark matter particle masses 
satisfy the criterion that M$_{\rm{core}}$/M$_{\rm{DM}} >$ 100.0 and we have found previously that 
these three particle masses (M$_{\rm{DM}}$ = 103 \msolarc, M$_{\rm{DM}}$ = 8 \msolar and 
M$_{\rm{DM}}$ = 0.6 \msolarc) produce converged results for this field strength 
(or equivalently baryonic core mass since the accretion rate depends on the core temperature 
which depends on the radiation field strength). \\
\indent In Figure \ref{HaloA:MassScaling_Mass} we have plotted the enclosed baryon mass scaled 
by the minimum dark matter particle mass against the enclosed mass ratio (i.e. enclosed 
dark matter mass divided by enclosed baryon mass). The representation clarifies that the scaling 
ratio must exceed a value of 100.0 at the point of baryon - dark matter enclosed mass equality 
(and the core must also remain baryon dominated) in order to fulfill our resolution requirements
as found previously. \\
\indent We therefore present this empirically derived result as a ``rule of thumb'' - 
\begin{equation}
\mathrm{M_{core}/M_{\rm{DM}} > 100.0}
\end{equation}
\emph{The minimum dark matter particle mass used must be a factor of one hundred less than the 
baryonic mass contained within a sphere in which the baryon mass dominates.} \\
However, we caution that this relationship does not provide a sufficient condition. It should 
instead be used as a starting point for conducting convergence tests for high redshift collapse 
simulations. Our simulations indicate that failing to adhere to this relationship will most likely 
produce spurious results and at the very least this relationship 
should be satisfied and convergence checked before performing full production runs.

\section{Conclusions} \label{Sec:Conclusions}

We have conducted a comprehensive study of the dark matter particle resolution requirements
necessary to study the collapse of gas at high redshift in the presence of a background radiation
field below the Lyman limit. We have used the publicly available \enzo code to conduct the study. 
\texttt{Enzo}, like most other grid (and particle) based codes, represents dark matter as discrete 
particles while solving the hydrodynamic equations on a Eulerian grid. In this light, we tested a 
number of different parameters that can potentially affect the interaction of the dark matter 
particles with the collapsing baryonic core. \\
\indent We investigated the effect of ``smoothing'' the dark matter particles onto the grid in 
regions of very high density in order to negate any two body scattering effects that may occur when 
the cell size becomes very small. We probed the effect of increasing the dark matter resolution 
around regions of interest by employing a particle splitting algorithm and tested its effectiveness. 
Furthermore, we varied the background radiation field from having no external radiation, J = 0
J$_{21}$, up to J = 500 J$_{21}$ where J$_{21}$ is in the usual units of $1 \times$ \J. \\
\indent We find that smoothing particles onto the grid provides some relief from the effects 
of spurious two body dark matter interactions and is most effective when the dark matter 
resolution is already comparatively high. With a dark matter particle mass of  
$\mathrm{M_{\rm{DM}} \lesssim 100}$ \msolarc, smoothing the particles onto the grid produces 
more converged results, when compared against the non-smoothed case, as the dark matter resolution is
increased but nonetheless having a high dark matter resolution was the crucial and overriding factor. 
At more intermediate particle masses,  $\mathrm{M_{\rm{DM}} \gtrsim 100}$ \msolarc, smoothing can 
alleviate some of the effects of two-body scattering (compared to the unsmoothed case) but it 
cannot negate them entirely and is somewhat case (halo) dependent. When the dark matter mass 
resolution is low ( i.e. $\mathrm{M_{\rm{DM}} \gtrsim 1000}$ \msolarc) smoothing provided no 
extra positive effect to alleviate spurious effects as expected. \\
\indent In summary, smoothing is most effective when the dark matter resolution is already high 
as it wipes out any lingering discreteness effects. It is significantly less effective as the mass 
resolution drops and it should not be seen as a fix for under resolved dark matter simulations. 
As a result we advocate the use of smoothing \emph{in conjunction with} high particle resolution 
only.   \\
\indent In examining the effectiveness of particle splitting we find very encouraging results.
We initially tested the effect of splitting on the properties of dark matter in dark matter only
simulations and found that the splitting added little or no additional numerical noise to the 
results finding that dark matter density profiles and enclosed mass profiles remained 
highly converged both before and after splitting. After adding in baryons, splitting 
the dark matter enabled us to remove spurious dark matter effects from otherwise
under-resolved simulations. \\
\indent Using an external radiation field with an energy below the 
Lyman limit we found that the particle mass required varied  with the strength of the 
radiation field. This is not surprising since a strong radiation field effectively dissociated
\molH (and its pathway element H$^{-}$) reducing the cooling ability of the halo thus increasing the 
accretion rate making it more massive. We found that as the halo grew the dark matter particle 
mass required also grew (i.e. less dark matter resolution was required). This means that when 
one is examining the affect that a very strong radiation field has on a collapsing halo the 
dark matter particle mass may be greater (by several factors), without any loss in numerical 
accuracy, compared to case where haloes are collapsing in the absence of an external radiation 
field or in the presence of a weak background field.\\
\indent We find that a useful ``rule of thumb'' is that the ratio of the enclosed baryonic mass 
within the core of the halo to the individual dark matter mass must exceed 100 i.e. 
${M_{\rm{core}} / M_{\rm{DM}}} > 100.0$.  We found that, for the simulations conducted here, this 
relationship must always be satisfied to produce converged results. We therefore advise that 
this relationship always be fulfilled and that appropriate convergence tests be conducted in 
parallel to ensure reliable results. \\
\indent While this study was carried out using the grid code \enzo we expect the results to hold 
for other grid and non-grid codes (e.g. SPH codes) that use the N-body technique to 
represent the dark matter component. On a final note it is appropriate to mention the recent 
work of \cite{Hahn_2015} who describe a method to follow the dark matter evolution using a 
six dimensional phase approach rather than the traditional N-body model studied here. Their
approach dispenses with the discrete nature of the N-body method and may help to significantly 
suppress any discreteness effects that can arise in modelling collisionless systems in the future.

\section*{Acknowledgements}
\noindent J.A.R. and P.H.J. acknowledge the support of the Magnus Ehrnrooth Foundation, the Research
Funds of the University of Helsinki and the Academy of Finland grant 274931.
J.H.W. acknowledges support by NSF grants AST-1211626 and
AST-1333360.  The numerical simulations were performed on facilities 
hosted by the CSC -IT Center for Science in Espoo, Finland, which are financed by the 
Finnish ministry of education. 
Computations described in this work were performed using the publicly-available \enzo code 
(http://enzo-project.org), which is the product of a collaborative effort of many independent 
scientists from numerous institutions around the world.  Their commitment to open science has 
helped make this work possible. The freely available astrophysical analysis 
code YT \citep{YT} was used to construct numerous plots within this paper. The authors would 
like to express their gratitude to Matt Turk et al. for an excellent software package. 
J.A.R. would also like to thank Greg Bryan for useful discussions leading to this work. Finally, we 
would like to thank the referee, Greg Bryan, for a very detailed report which greatly enhanced the 
final manuscript.

\appendixtitleon
\appendixtitletocon
\begin{appendices}
\section{Running at the maximum refinement level} 
\label{appendixA}
\begin{figure*}
  \centering 
  \begin{minipage}{175mm}      \begin{center}
      \centerline{
        \includegraphics[width=9cm]{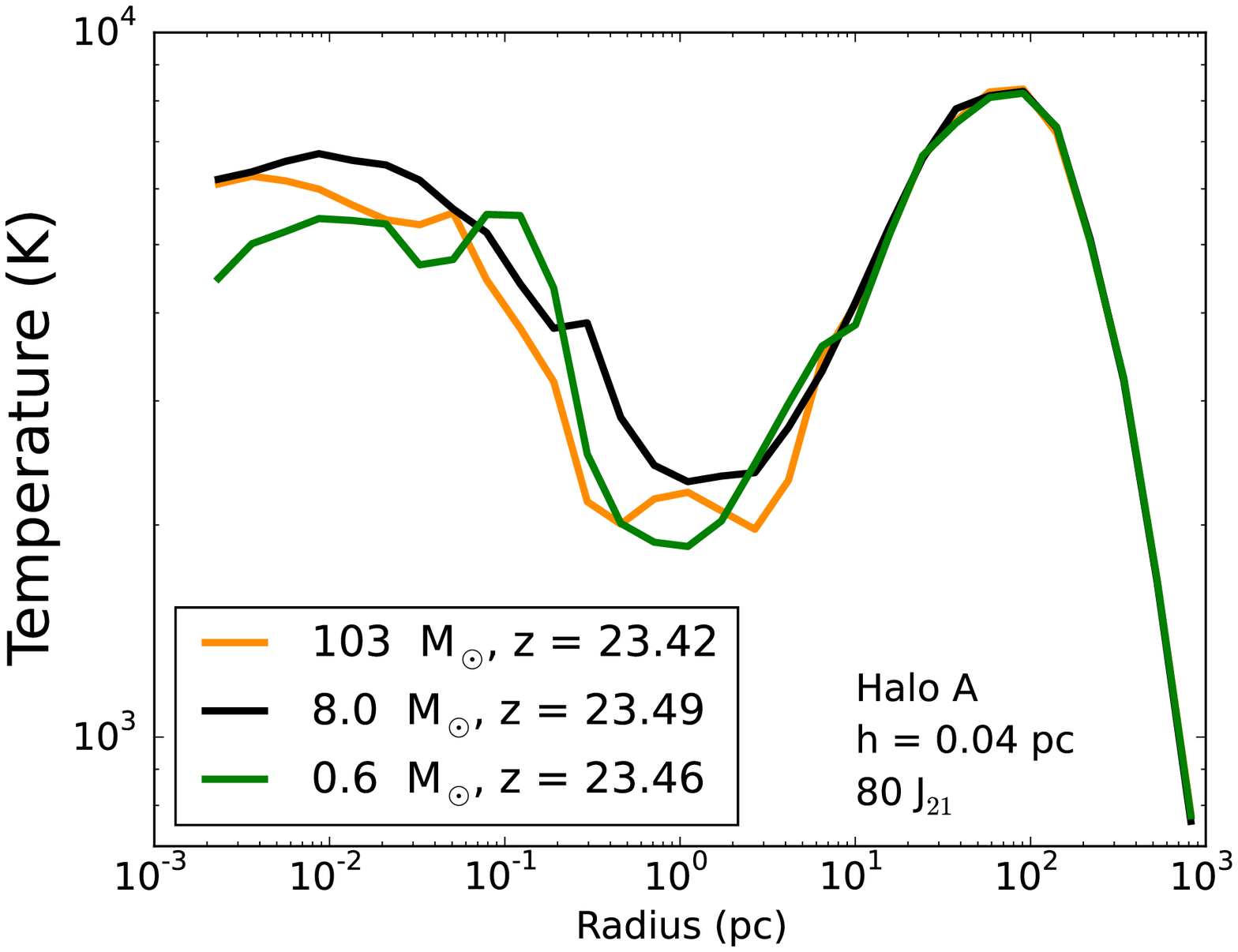}
        \includegraphics[width=9cm]{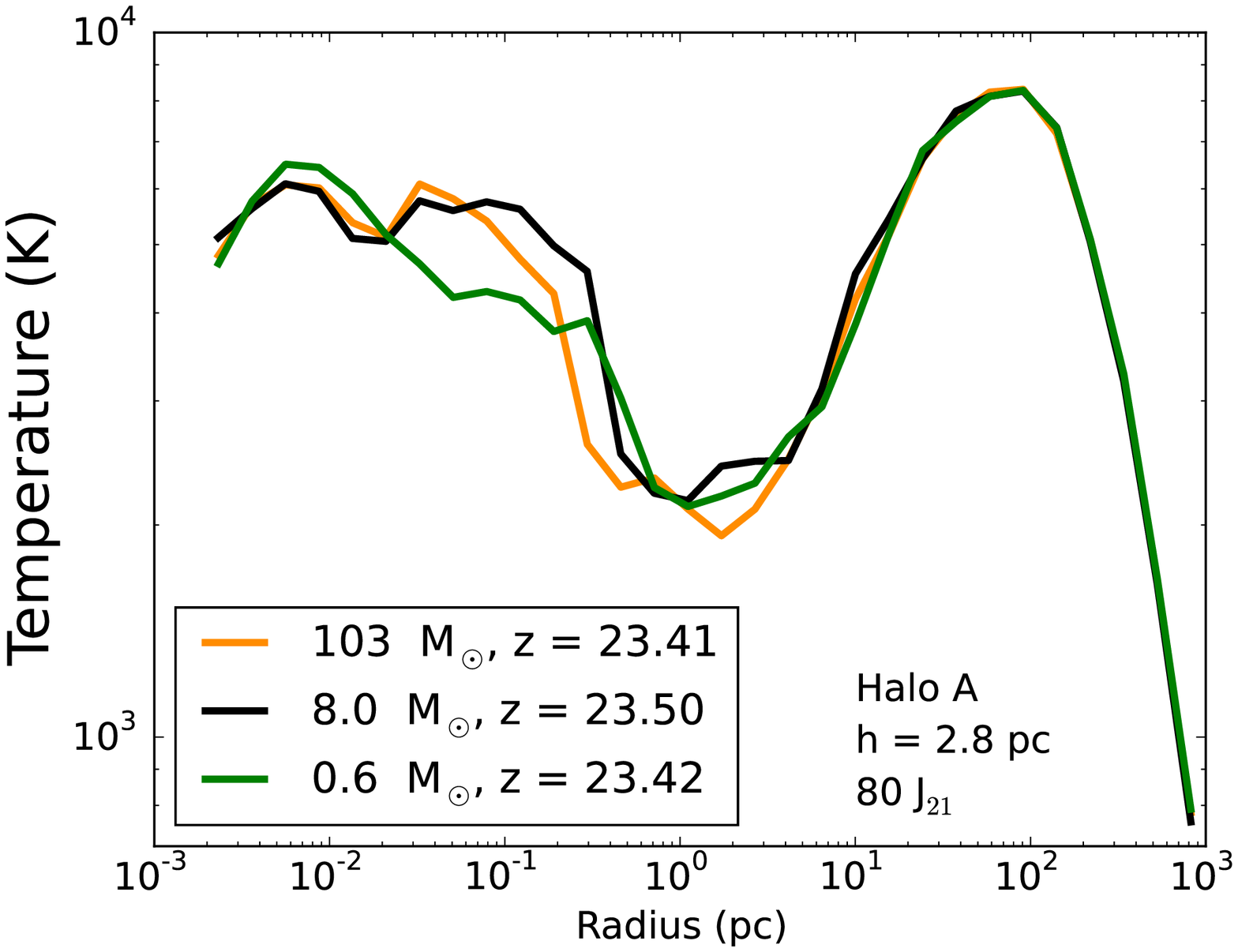}}
      \caption[]
      {\label{HaloA:Extra}
        \textit{Halo A}: The left hand panel shows the spherically averaged temperature profile
        in the centre of the collapsing halo a short time ($\sim$ 130,000 years) after running at 
        the maximum refinement level. No smoothing of the dark matter particles occurs in the left 
        hand panel. The right hand panel shows the same temperature profile for simulations in which 
        smoothing of the dark matter particles occurs at 2.8 comoving pc (refinement level 12). 
      }
    \end{center} \end{minipage}
\end{figure*}

\begin{figure*}
  \begin{minipage}{175mm}      \begin{center}
      \includegraphics[width=16cm]{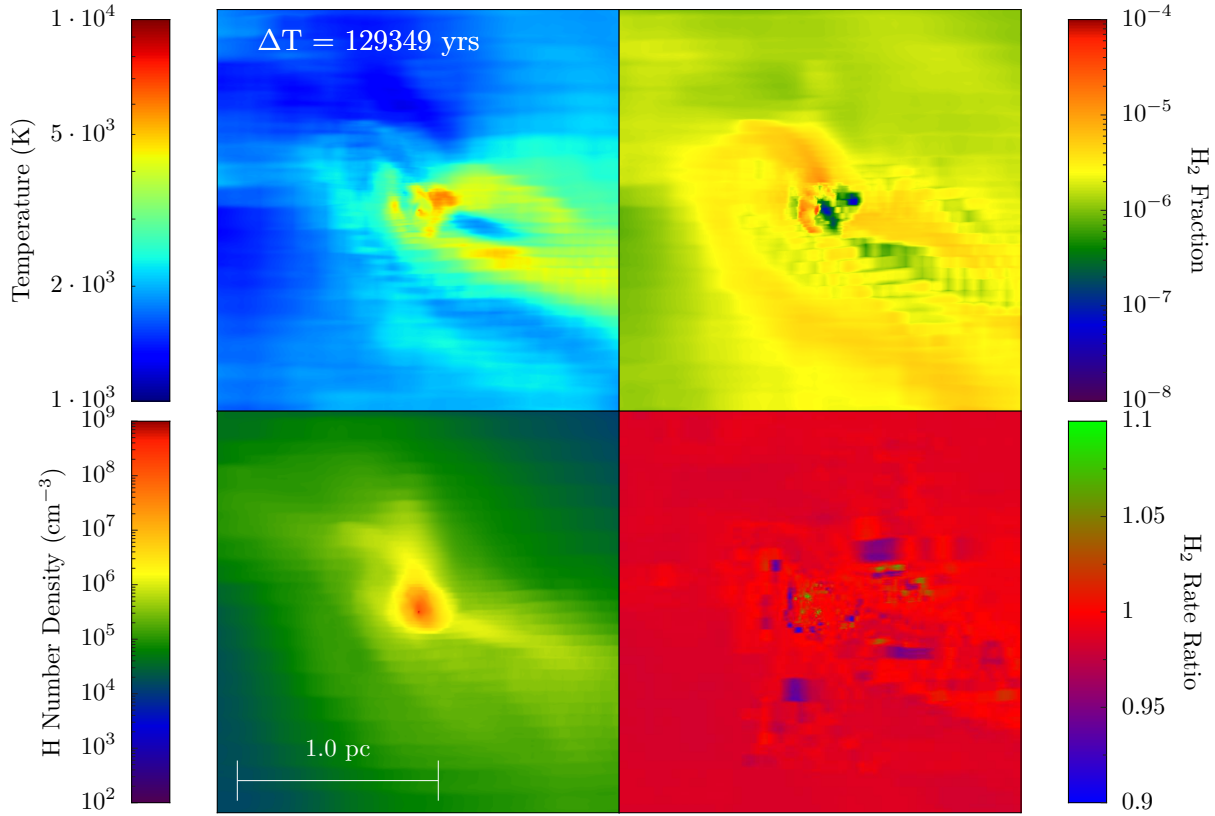}
      \caption[]
      {\label{HaloA:ExtraVis}
        \textit{Halo A}: The four panel projection shows the state of the gas within the core of the 
        halo approximately 130,000 years after reaching the maximum refinement level. Each 
        projection is aligned perpendicular to the angular momentum vector.
        In the top left panel is shown the temperature projection showing the central ``hot'' 
        core surrounded by cooler gas. The \molH fraction is shown in the top right panel 
        and shows that the core contains clumps of molecular gas surrounded by gas where 
        the \molH fraction is significantly lower. The lower right panel shows the ratio 
        of the \molH formation rate to the \molH dissociation rate. Green areas represent 
        areas where \molH is forming while blue areas represent areas where the \molH is being 
        dissociated. Finally, in the lower left panel the H number density is shown, in the 
        core of the halo the H number density is at its maximum.
}
 \end{center} \end{minipage}
\end{figure*}

\indent As discussed in \S \ref{Sec:Smoothing}, Figure \ref{HaloA:Extra} shows the 
temperature profile for Halo A for 
both a non-smoothed (i.e. h = 0.04 pc, left panel) and a smoothed run (i.e. h = 2.8 pc, 
right panel) further into its evolution than presented in Figure \ref{HaloA:MultiPlot_J80}.
The simulations are run at the maximum refinement level to examine the 
further evolution of the core. As mentioned previously, running at the maximum refinement 
level artifically prevents collapse while the Jeans length is also not resolved. This may 
potentially induce unphysical results, however we are careful, as detailed below, to only 
run the simulation for a few timesteps. In this case we plot only the results for the runs in 
which the collapsing core is gas dominated and therefore exclude simulations S6646 and S830 
from both plots so as to exclude any potential dark matter contamination effects which may 
affect the physical results. \\
\indent The simulations have been restarted from the last output and allowed to run at the 
maximum refinement level for a short time ($\sim$ 130,000 years). No artificial pressure floor is 
employed here so as to exclude any artificial heating effects although in principle running at 
the maximum refinement level without any artificial pressure support can lead to numerical 
instabilities \citep{Machacek_2001}. We are therefore careful to run the simulations only for a 
few timesteps at this point. As can clearly be seen, as the simulation is allowed to progress, 
the temperature in the core increases and the core heats up to close to $\rm{T} \sim 6000$ K in 
all cases. The reason for this heating is that dense clumps of neutral hydrogen form in the core 
promoting (collisional) dissociation of \molH through the reaction 
\begin{equation}
\rm{H_2 + H \rightarrow 3H}
\end{equation}
We have used the formulation given in \cite{Martin_1996} to model this reaction.
The dissociation rate then exceeds the \molH formation rate in regions within the core of the halo
but conversely the formation rate exceeds the dissociating rate in other neighbouring parts 
resulting in the formation of hot and cold spots within the core of the halo. The formation rate 
is controlled, at these densities, by the reaction
\begin{equation}
\rm{H^- + H \rightarrow H_2 + e}
\end{equation}
In Figure \ref{HaloA:ExtraVis} we have plotted four projection plots illustrating this effect. 
The projections are made approximately 130,000 years after reaching the maximum refinement level 
using a S103 (with no-smoothing) output for illustration (although S8 or S06 could also have been 
used). In the bottom left panel we show the neutral hydrogen number density 
showing the formation of a dense core, in the top left panel we show the temperature projection 
clearly displaying regions of hot and cold gas within the core. The \molH fraction is shown in 
the top right panel, reflecting the temperature 
projection, we see regions of low \molH fraction surrounded by regions of high \molH fraction which 
dramatically affects the cooling in these regions and causes bumps in the temperature profile as 
shown in Figure \ref{HaloA:Extra}. Finally, in the bottom right panel we show the ratio of the 
\molH formation rate to the \molH dissociation rate. Again, we see regions of formation and 
dissociation within the core consistent with the other projections. The core of the halo 
is essentially sitting on a knife edge between \molH formation and dissociation. \\
\indent It should also be noted that our simulations neglect the potentially important effects of 
self-shielding. However, a study of the complete effects of self-shielding on the core 
hydrodynamics is beyond the scope of this paper.

\end{appendices}
\end{document}